# Chemical Differentiation of Planets: A Core Issue

Hervé Toulhoat[1] and Viacheslav Zgonnik[2]
[1] Sorbonne Université, UPMC, CNRS, Laboratoire de Réactivité de Surface, 4 Place Jussieu, F-75005, Paris, France; herve.toulhoat@orange.fr
[2] Natural Hydrogen Energy LLC, French Branch: 31 Rue Raymond Queneau, F-92500 Rueil Malmaison, France
*Received 2021 May 27; revised 2021 September 9; accepted 2021 October 12; published 2022 January 14*

## Abstract

By plotting empirical chemical element abundances on Earth relative to the Sun and normalized to silicon versus their first ionization potentials, we confirm the existence of a correlation reported earlier. To explain this, we develop a model based on principles of statistical physics that predicts differentiated relative abundances for any planetary body in a solar system as a function of its orbital distance. This simple model is successfully tested against available chemical composition data from CI chondrites and surface compositional data of Mars, Earth, the Moon, Venus, and Mercury. We show, moreover, that deviations from the proposed law for a given planet correspond to later surface segregation of elements driven both by gravity and chemical reactions. We thus provide a new picture for the distribution of elements in the solar system and inside planets, with important consequences for their chemical composition. Particularly, a 4 wt% initial hydrogen content is predicted for bulk early Earth. This converges with other works suggesting that the interior of the Earth could be enriched with hydrogen.

## 1. Introduction

According to the most widely accepted model, the solar system formed from the gravitational collapse of a fragment of a giant molecular cloud (Brahic 2006). The accretion of the gas and dust, left over from the Sun's formation, forms planetesimals that collide and aggregate into larger bodies: planets. In this model, the chemical compositions of the planets depend on temperature. In the inner solar system, which was supposedly warm, terrestrial planets form from compounds with high melting points, while more volatile compounds accumulate beyond the so-called "ice line," where they condense to form giant planets (McSween & Huss 2010). The elemental composition of the most primitive accreting material before condensation is supposed to be similar to the CI chondrites (carbonaceous chondrite, Ivuna-type), meteorites that are considered the least chemically fractionated when relative abundances are compared to the photosphere (McSween & Huss 2010). Notwithstanding, inner planets, including Earth, are distinct from any type of extant primitive meteorites or their mixtures (Drake & Righter 2002; McSween & Huss 2010; Campbell & O'Neill 2012). Heterogeneous and homogeneous accretion models, together with coherent radial differentiation models integrating geophysical properties of planets, are debated in order to explain observed elemental compositions of planets (Montmerle et al. 2006). Existing accretion and differentiation models are shown to be insufficient to explain the various elemental compositions of planetary materials (Bertka 1998), and more specifically the Earth's (Javoy 1999).

Therefore, new approaches are necessary to understand chemical differentiation processes in the solar system. A few decades ago, V. Larin demonstrated that there is a correlation between the abundance of elements on the Earth relative to the Sun and their first ionization potential (IP; here and below, by ionization potential, we mean first ionization potential) (Larin 1973). To explain this apparent correlation, he used space plasma physics-based cosmological hypotheses developed by Hoyle (1960) and by Alfvén & Arrhenius (1976). They proposed that at some moment of time, a portion of matter in the solar system was ionized and was moving perpendicularly to the magnetic field lines of the proto-Sun. This process caused the separation of elements by their ionization potential, as elements with lower IPs were captured by Lorentz forces in the first place. In the same, coupling between the magnetic field and the ionized matter in the disk transferred the angular momentum from the proto-Sun to the accretion disk. Later, Shu et al. (2007) and Mohanty & Shu (2008) proposed that a similar role might have been played by sub-Keplerian rotation of the disk of accreting, partially ionized gases subjected to a magnetic field dragged from interstellar space by the overall solar system forming gravitational collapse. The relationship between IP and the chemical composition of Earth and the Moon was also examined by Hauge (1971) with the goal to test Alfvén's theory of the evolution of the solar system which combines electromagnetic forces with gravitational forces (Alfvén 1954). However, without normalizing elements' abundances to one element (usually silicon is taken as reference), Hauge could not extract a convincing evidence from the data.

Since the time of these proposals, the understanding of the early evolution of circumstellar disks has advanced significantly, especially due to many observational discoveries. For instance, we now know that all young solar-like stars emit X-rays and that such irradiation ionizes the matter in accretion disks (Montmerle et al. 2006). Even weakly ionized matter will "stick" to the magnetic field, and if the matter is moving, this could eventually lead to differentiation of elements depending on their IPs. However, this idea was not further subjected to theoretical analysis.

Herein, we reappraise the available data for the inner solar system composition and confirm that there is a correlation between the IP of elements and their distribution in the planetary bodies. We explain it from basic concepts of statistical physics, which results in a new predictive model for the distribution of elements in the solar system.

## 2. Theory of Magnetically Driven Chemical Differentiation across the Solar System

The differentiation factor of an element for a planetary body is defined as the ratio of its relative abundance in this body to its relative abundance in the solar system with silicon abundances taken as references. We have computed differentiation factors for Earth using Earth's uppermost continental crust data (Lide 2005), corrected to include major elements in the hydrosphere (H, O, Cl, Mg, Na, Ca), and for the solar system we take the recommended spectroscopically measured relative abundances in the solar photosphere (Lodders 2003) as the composition of the Sun has not change since it was formed (with a few exceptions mentioned in the Appendix).

For our analysis, we use only the empirical data, i.e., obtained by measurements, and we do not use any data that are based on models or assumptions. For Earth, we only have measurements for the material of its outer geosphere. We do not have any samples from the Earth's inner mantle or core and rare samples from the mantle (e.g., kimberlites) are not representative. Therefore, we use only averaged data from the uppermost continental crust. We justify our choice of continental crust by the fact that it is representative of the oldest and most conserved part of the lithosphere and according to some estimates (Peterson & Depaolo 2007) represents almost 80% of the total mass of the crust. In the SI we compare the Earth's uppermost continental crust data we used for our work with data for actual samples of oceanic crust rocks and demonstrate that the mass fractions normalized to Si are indistinguishable. Therefore, we can use only the uppermost continental crust data.

Certainly, the crust has been affected by all the different processes during the history of the Earth. However, for the uppermost continental crust, the sampling is representative from all continents and exhaustive enough to erase local fluctuations of composition resulting from relatively superficial geological processes. Earth's crust data from the CRC Handbook (Lide 2005) are consistent with other sources available (Turekian 1970; Shaw et al. 1976; Wänke et al. 1984; Weaver & Tamey 1984; Taylor & McLennan 1985; Earnshaw & Greenwood 1997). The data represent averages of many actual analytical measurements and therefore are statistically significant. Several exclusions and corrections for radiogenic nuclides are addressed in the Appendix.

The correlation between the differentiation factor and ionization potential of elements is shown in Figure 1 for elements from H to U. The semilog plot exhibits an average slope of $-1.175\,\mathrm{eV}^{-1}$, with a squared coefficient of correlation of 0.59 (Figure 1, inset). The physical meaning of this correlation is shown below to be linked to electromagnetic forces, while observed outliers are linked to radial differentiation as demonstrated in Section 4. By contrast, we show in the Appendix A (Figures A4 and A5) that temperatures of condensation of 50% of the mass into the most stable mineral, 50% $Tc$, are very poorly correlated to differentiation factors ($R^2 = 0.4$ and 0.2 with and without noble gases, respectively) as well as to IP.

Since the abscissa-axis in this plot is an energy scale, the law is reminiscent of a Boltzmann distribution. We understand this as follows: atoms in the protoplanetary gaseous accretion disk are flowing toward the proto-Sun or beyond some escape distance from its gravitational attraction, away because of its rotational instability. This determines a net radial flux of matter. However, a fraction of this matter becomes ionized by radiation mostly emitted by the proto-Sun (including, for instance, those resulting from radioactive decay of short-lived isotopes, or merely X-rays resulting from a very high surface temperature) and then is diverted toward equilibrium orbits by the centripetal Lorentz force exerted by the magnetic field normal to the nebular disk. Viewing the ionization potential IP($M$) of a given element $M$ as the activation energy for its ionization, the molar (or mass) fraction of $M$ $\left(\frac{X^+}{X_{SS}}(M)\right)$ trapped in orbit at average distance $d$ from the proto-Sun is proportional to its ionization probability:

$$\left(\frac{X^+}{X_{SS}}(M)\right) = \exp\left(\frac{-\mathrm{IP}(M)}{k_B T_{elG}(d)}\right) \quad (1)$$

where $X_{SS}(M)$ is the initial average abundance of element $M$ in the solar system. Here, we define $T_{elG}(d)$ as the local electronic temperature of the plasma depending on the distance from the ionizing source.

We consider the protoplanetary gas as a dilute atomic nonthermal plasma absorbing radiation from the proto-Sun and emitting toward the cosmic background. Locally (at distance $d$ from the proto-Sun) and on a microscopic space and timescale, the absorption (or emission) spectrum will exhibit strong lines corresponding to the electronic transitions for ionizations in the range of energy ∼4–25 eV corresponding to photons able to trigger the first ionization of the chemical elements.

In that range of energy the power absorbed per unit volume of gas is therefore strictly proportional to the flux of photons from the proto-Sun, itself proportional to $\Omega(d)$, the local solid angle of observation of the proto-sun:

$$\Omega(d) = \pi\left(\frac{R_{PS}}{d}\right)^2. \quad (2)$$

With $R_{PS}$ the radius of the proto-Sun. Once a stationary regime has been reached, the average local kinetic energy of the gas is locally equal to the average local input of radiative energy so that:

$$k_B T_G(d) = k_B T_{elG}(d)\pi\left(\frac{R_{PS}}{d}\right)^2 \quad (3)$$

with $T_G(d)$ the local protoplanetary gas temperature. We have considered that any finite local optical density will equally affect the input and output of energy from any elementary volume of gas, and therefore cancel out in the balance.

Besides, macroscopically, that is, at astronomical space and timescales, the protoplanetary plasma can be considered to be in equilibrium with the cosmic background so that at distances far enough from the proto-Sun:

$$T_G(d) \approx T_{CB} \quad (4)$$

where $T_{CB}$ is the cosmic-background temperature, and thus the lower limit for gas temperature in a protoplanetary disk or in the interstellar medium. We can therefore propose the following ansatz, with the distance to Sun $d$ in au:

$$\frac{T_G(d)}{T_{CB}} = \max\left\{T_G(1\ \mathrm{au})\left(\frac{1}{d}\right)^\alpha, 1\right\}. \quad (5)$$

We will show below that the exponent $\alpha$ can be consistently determined for our solar system from geochemical data collected by space probes for solar system bodies.

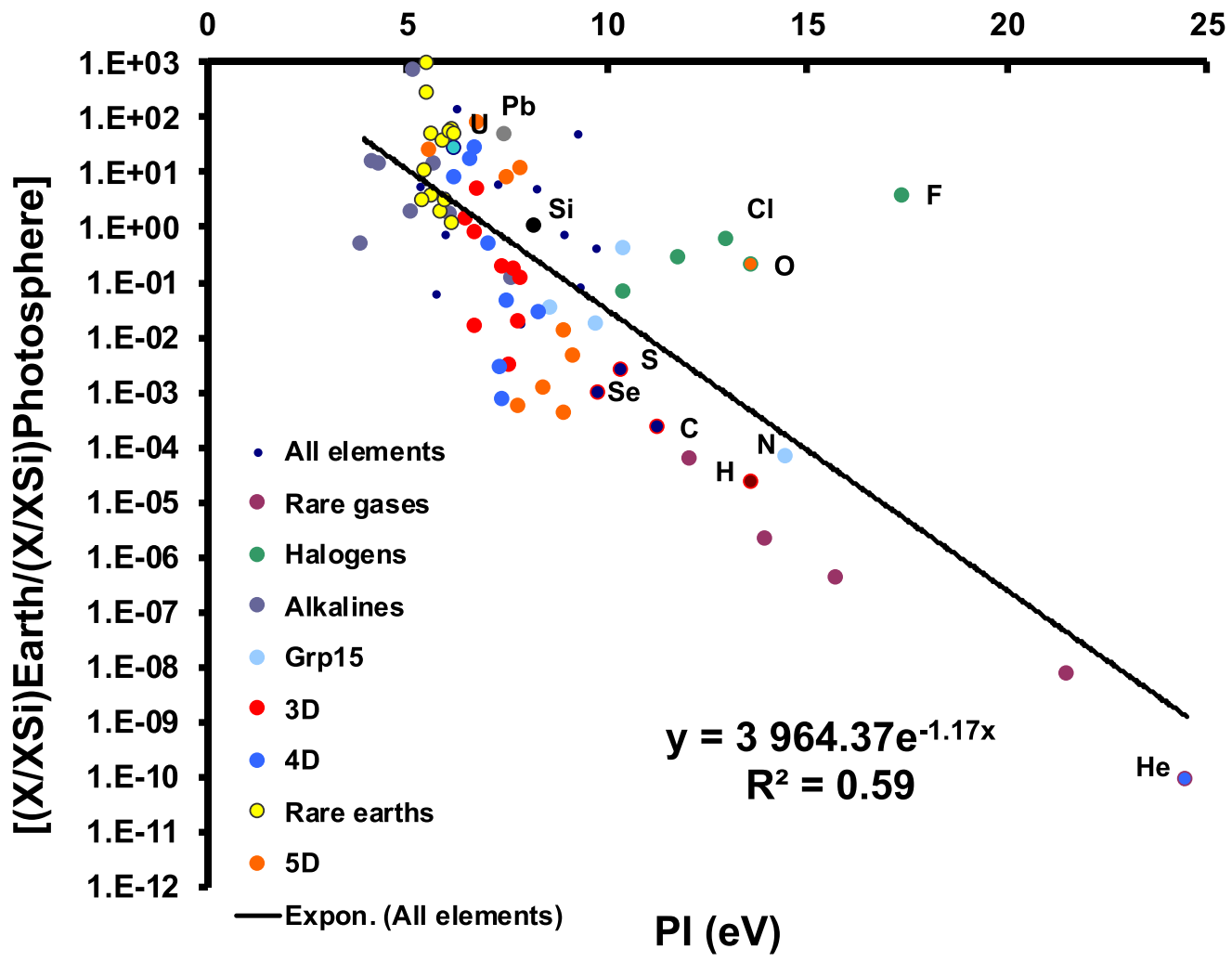

**Figure 1.** Earth crust differentiation factors vs. first ionization potential for elements H to U. The slope of the regression line in the semilog plot, excluding He, is $-1.175\ \mathrm{eV}^{-1}$.

According to Equation (3), for a dilute atomic cloud irradiated by a spectrum of ionizing radiation, thermalization is mainly achieved by ionizing absorption and de-ionizing emission. The ionizing flux decreases with the solid angle of sight as distance $d$ from the source increases. Thermal equilibration is obtained by keeping a fraction of the elements in the ionized state, which decreases exponentially as IP increases. At close distance from the proto-Sun (e.g., $d \ll 1$ au), this fraction is significant for the lower IPs ($\sim$4 eV) and decreases exponentially with increasing IP. It decreases rapidly with increasing distance $d$ as $T_G(d)$ and $T_{elG}(d)$ decrease, but because dilute atomic gas in space cannot be cooler than the cosmic-background temperature, Equation (3) implies that $T_{elG}(d)$ increases quadratically with $d$ above some distance at which it passes a minimum. Therefore, for very large distances, the ratio of ionized over neutral atoms approaches 1, even for the elements with the higher IPs, although the flux of highly energetic photons received by any atom from the hot central body decreases as $d^{-2}$. This apparent paradox can be solved considering that if the ionization probability of an atom per unit time decreases, being proportional to the flux of photons, the timescale to reach thermal equilibrium with the cosmic background, i.e., heat the cold and dilute ions up to $T_{\mathrm{CB}}$ by transfer of energy from the hot electrons, may increase indefinitely. The cold plasma state of the remote protoplanetary cloud can therefore be seen as an energy storage enabling us to satisfy everywhere the condition of thermal equilibrium with the cosmic background.

Finally, considering the normalization of abundances with respect to silicon, and assuming the gravitational aggregation of a planetary body at distance $d$, the law of magnetic chemical differentiation of planets is written as:

$$\frac{\left(\frac{X}{X_{\mathrm{Si}}}\right)}{\left(\frac{X}{X_{\mathrm{Si}}}\right)_{\mathrm{SS}}}(M, d) = e^{\left(\frac{-[\mathrm{IP}(M)-\mathrm{IP(Si)}]}{k_B T_G(d)}\pi\left(\frac{R_{\mathrm{PS}}}{d}\right)^2\right)} = f_V\ (M, d). \quad (6)$$

Where $f_V(M, d)$ is the bulk differentiation factor for element $M$ of a planetary body $V$ gravitating at average distance $d$, and the subscript solar system (SS) refers to the average solar system. In the case of the Earth ($d = 1$ au), we notice that the regression value obtained for elements H to U, excluding He, (Figure 1), $-1.175\ \mathrm{eV}^{-1}$ can be almost recovered from Equation (6) by setting $R_{\mathrm{PS}} = 2R_S = 0.00928$ au, with $R_S$ representing the present solar radius, and $T_G$ ($d = 1$ au) $= T_{\mathrm{CBP}} = 2.75$ K, with $T_{\mathrm{CBP}}$ the present cosmic-background temperature (the latter is in principle lower than it was at the time of solar system formation 4.5 Gy ago, in view of the expansion of the universe). These values are plausible although arbitrary. However, provided Equation (6) holds, one expects linear correlations between logs of experimental differentiation factors of the elements for different planets, e.g., referred to Earth for which the most complete and precise set of relative abundance data are available; one may test our ansatz Equation (5), i.e., the $d^{-\alpha}$ scaling and the ratios $T_G(d)/T_G(1)$, indeed:

$$\lambda(M, d_1, d_2) = \frac{\ln(f_V\ (M, d_2))}{\ln(f_V\ (M, d_1))} = \frac{T_G(d_1)}{T_G(d_2)}\left(\frac{d_1}{d_2}\right)^2. \quad (7)$$

Therefore, $\lambda(d_1, d_2)$ is independent of elements. So

$$\lambda(1, d_2) = \frac{T_G(1)}{T_G(d_2)}(d_2)^{-2}. \quad (8)$$

According to our model, for planets beyond the asteroid belt, elemental compositions will hardly be distinguished from the Sun's photosphere composition, consistent with current estimations. Indeed, atoms cannot be separated by mass while in freefall in the gravitational field of the protostar at long range, but start "feeling" separation by electromagnetic forces akin to dipole–dipole interactions at shorter range. As mentioned above, since a dilute atomic gas in space cannot be cooler than the cosmic-background temperature, Equation (3) implies that $T_{elG}(d)$ increases quadratically with $d$ above some distance which happens, as shown by Figure 7, to be $d \approx 1$ au. Equation (1) then implies that all elements will become fully ionized at a large distance from the proto-Sun, while this translates from Equation (6) into differentiation factors approaching 1 for all elements, i.e., all atoms stay trapped in orbit by Lorenz forces able to sustain gravitational attraction. This is confirmed by our analysis in Figure 7 of correlations between available analytical data for planet surfaces presented in Figures 2–6. A further implication is that a large fraction of protostellar disk matter will remain in the outer regions undifferentiated from the primitive stellar matter and condense later into giant planets extremely rich in hydrogen. This is indeed the case for our solar system, with Jupiter, Saturn, Uranus, Neptune, and their satellites. Therefore, our model gives an explanation of the differentiation of planets into terrestrials close to the star, and gaseous giants far from the star.

## 3. Test of the Theory against Available Compositional Data of Bodies from the Solar System

We have tested Equation (6) using chemical analysis data available for asteroid belts and for rocky planetary bodies: Mercury, Venus, Mars, and the Moon. The numerical values we used are tabulated in Table A3 in the Appendix. The results are presented in Figures 2–6 for the asteroid belt, Mercury, Venus, Mars and the Moon, respectively. Each figure presents the correlation between the logs of differentiation factors of the planet and those of the Earth.

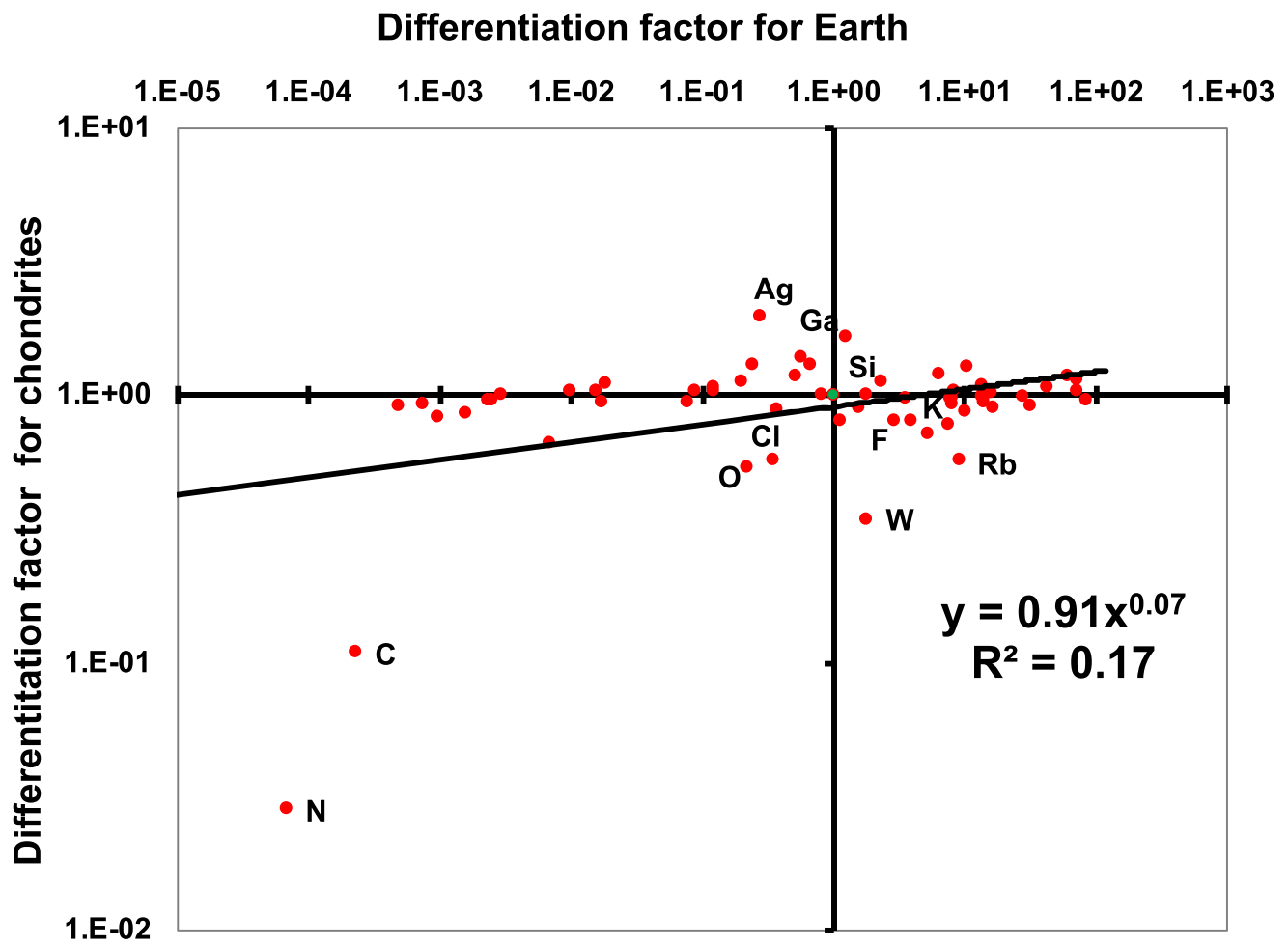

Figure 2. Correlation between the differentiation factors of the asteroid belt and the Earth (note that differentiation factors are available for planetary crusts only).

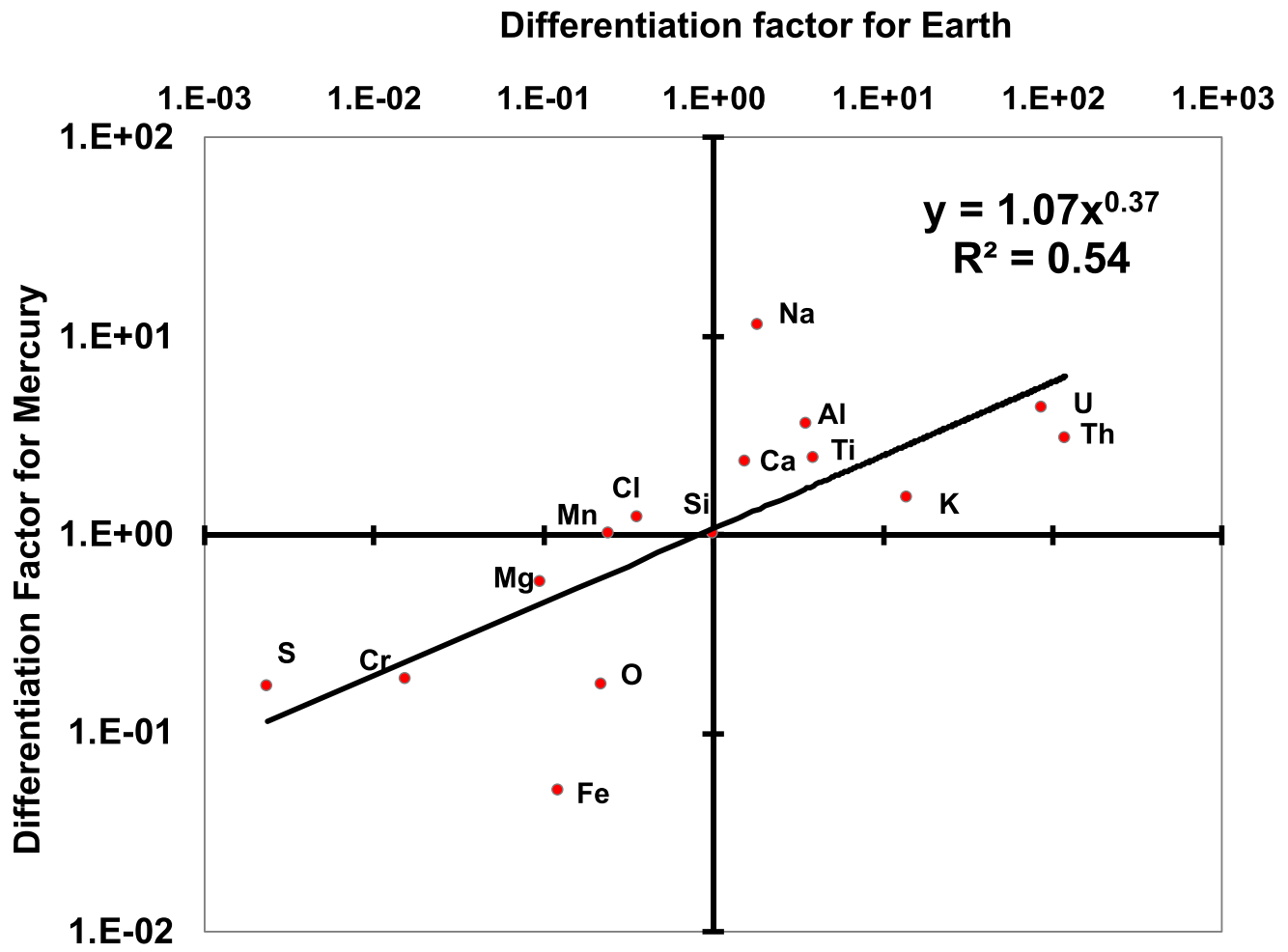

Figure 3. Correlation between the differentiation factors of Mercury and the Earth (note that differentiation factors are available for planetary crusts only).

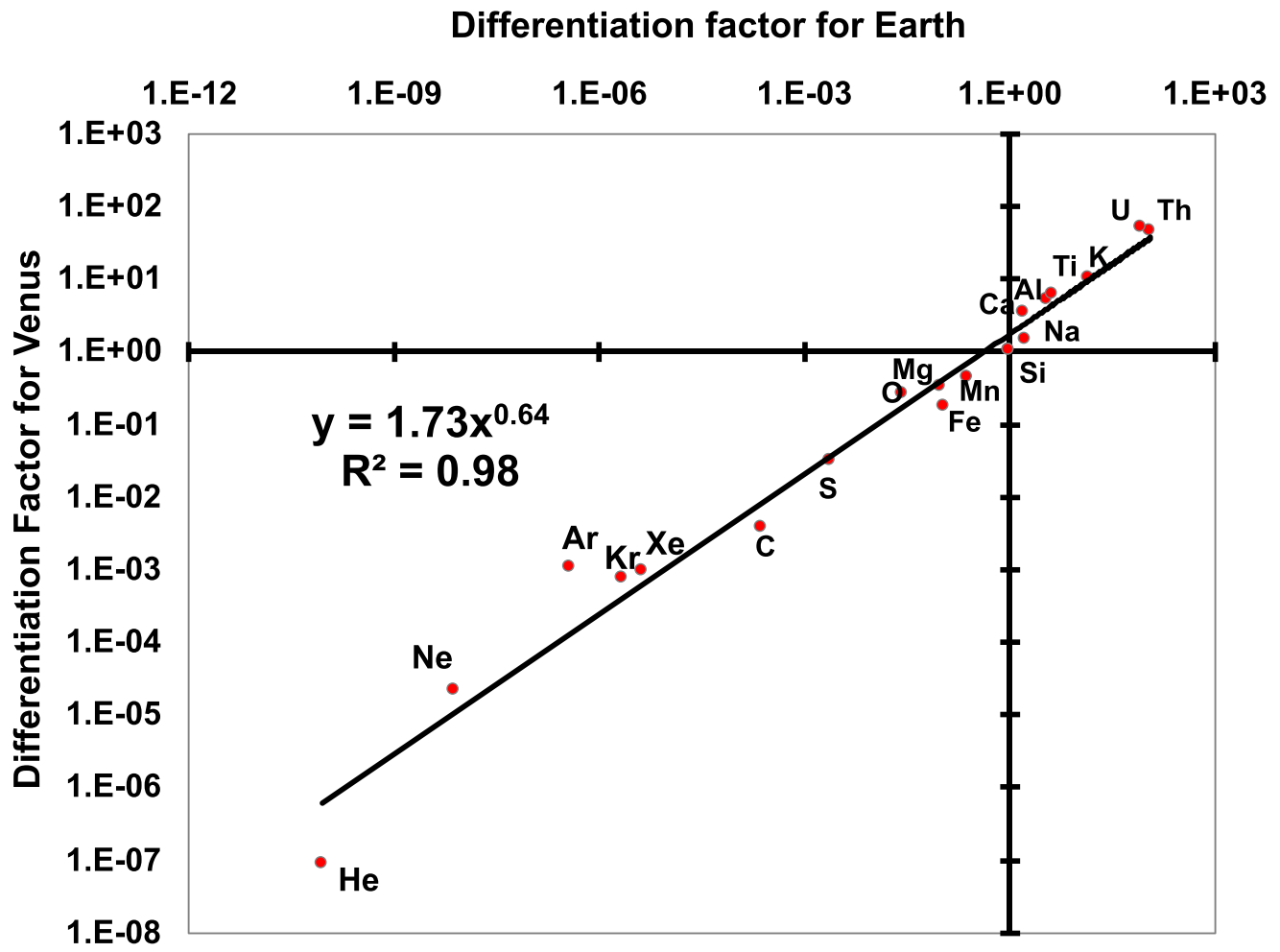

Figure 4. Correlation between the differentiation factors of Venus and the Earth (note that differentiation factors are available for planetary crusts only).

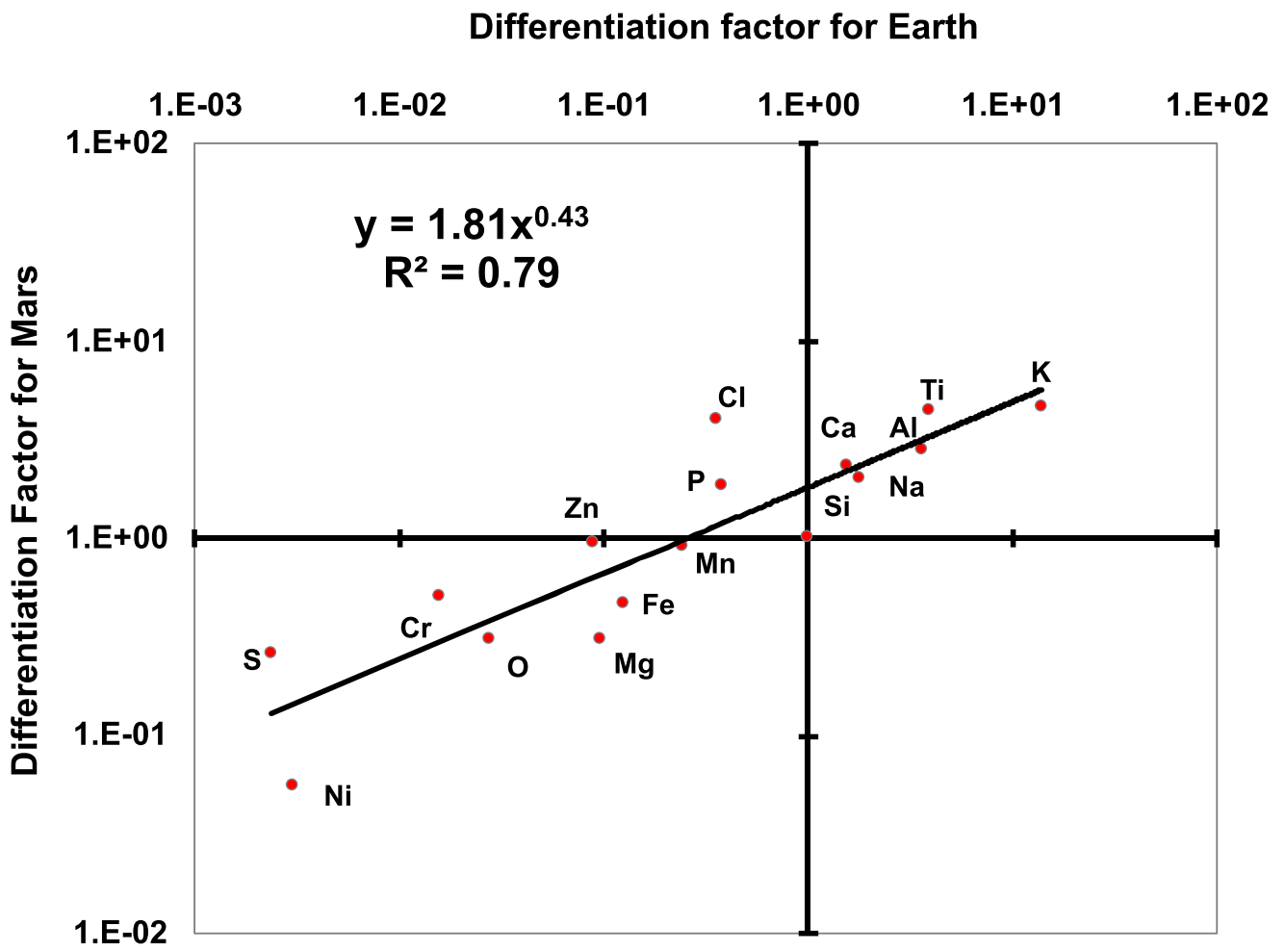

Figure 5. Correlation between the differentiation factors of Mars and the Earth (note that differentiation factors are available for planetary crusts only).

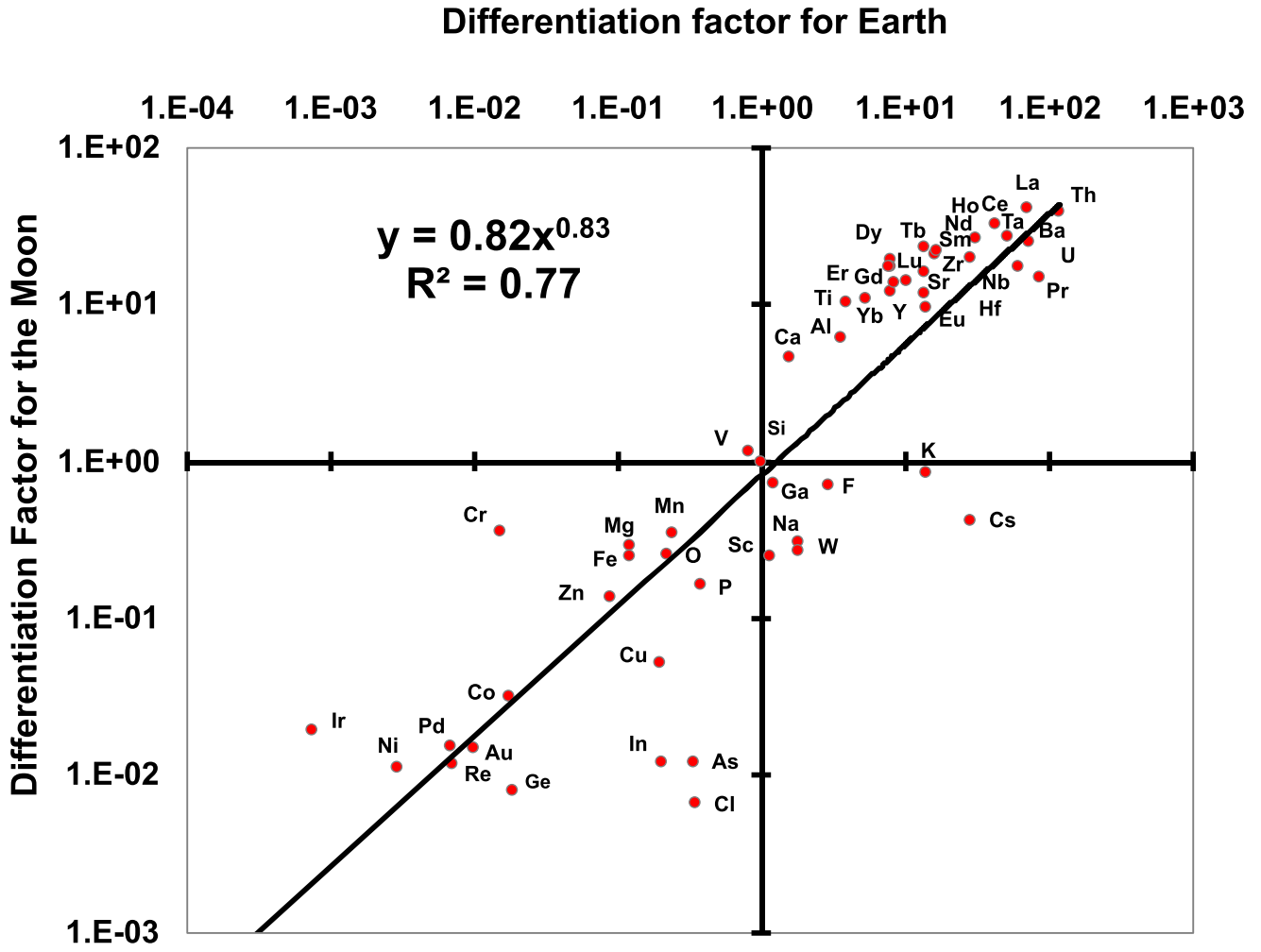

Figure 6. Correlation between the differentiation factors of the Moon and the Earth (note that differentiation factors are available for planetary crusts only).

Considering an average distance to the Sun of 3 au for the asteroid belt, we use elemental abundances in CI chondritic meteorites, which are supposedly derived from the most common C-type asteroids of the asteroid belt (Scott & Krot 2007) but also representatives of the primitive material of the early solar system. We exclude noble gases and hydrogen, which we consider to be strongly degassed from chondritic meteorites. Indeed, relative abundances of noble gases are strongly depleted from carbonaceous Ivuna- (CI) type meteorites in comparison to the solar photosphere, much more than are C and N (Pepin 2006). The correlation in Figure 2 is weak but provides a slope 0.0674, of the order expected for the large distance to the Sun compared to that of the Earth.

The inner planets are not believed to have migrated significantly over the age of the solar system. Therefore, in what follows, we assign distance to the Sun for planets as the average of their perihelion and aphelion, namely $d = 0.387$, 0.723, 1.524 au for Mercury, Venus, and Mars, respectively.

The composition of Mercury's surface was taken from the data from the MESSENGER mission (Nittler et al. 2011; Peplowski et al. 2011). The data are based on surface spectroscopy measurements (MESSENGER gamma-ray spectrometer), the only direct analyses available for this planet so far. Figure 3 demonstrates a moderate correlation ($R^2 = 0.54$) with slope of 0.3698.

Compositions of Venus rock samples from the Venera 13, 14, and Vega 2 missions (Abdrakhimov & Basilevsky 2002) were completed by the C, Ne, $^{36}$Ar, $^{84}$Kr, and $^{132}$Xe from Venus' atmosphere (Fegley 2007). Oxygen is deduced by mass balance over major elements. Figure 4 demonstrates an outstanding correlation ($R^2 = 0.975$) with a slope of 0.643.

Mars' crust composition was calculated after data from the Pathfinder, Opportunity, Spirit, and Curiosity missions (Foley 2003; Hahn 2009; Arvidson et al. 2010; McLennan et al. 2014; Schmidt et al. 2014) excluding analysis of soils, which can be affected by meteoritic dust. Oxygen is deduced by mass balance over major elements. Mars' atmosphere is strongly depleted in noble gases, compared to Venus and the Earth because of escape under the low gravity and low magnetic field. Consequently, as for chondrites, we have not included the relative abundances of those reported noble gases in our analysis. Figure 5 demonstrates a quite good correlation ($R^2 = 0.79$) with a slope of 0.434.

Figure 6 was built after experimental data from the Apollo 15, 16, and 17 missions to the Moon (Wanke et al. 1973). For each element, the mass fraction is averaged after all stations for all missions. It demonstrates a quite good correlation ($R^2 = 0.77$) with slope 0.83. This result is consistent with Moon elements inherited from gaseous protoplanetary matter in close vicinity of the Earth in agreement with other models (Ringwood 1986).

Summarizing, Figures 3–5 present moderate to very good linear correlations based exclusively on relative abundances gathered from experimental geochemical data. The slopes $\lambda$ in these correlations are therefore model-independent observations. From Mercury to Earth, we note the striking linear scaling of these figures with distance to the Sun, as shown in Figure 7. Moreover, beyond Earth, $\lambda$ scales as $\left(\frac{1}{d}\right)^2$, including the Moon, Mars, and the asteroids in the correlation. Then, according to our model (Equation (8)), we have for $d < 1$ au:

$$\lambda(d) = \frac{T_G(1)}{T_G(d)}\left(\frac{1}{d}\right)^2 = d, \text{ therefore:}$$

$$T_G(d) = T_G(1)\left(\frac{1}{d}\right)^3. \tag{9}$$

And for $d \geqslant 1$ au:

$$\lambda(d) = \frac{T_G(1)}{T_G(d)}\left(\frac{1}{d}\right)^2 = \left(\frac{1}{d}\right)^2, \text{ therefore:}$$

$$T_G(d) = T_G(1). \tag{10}$$

From our argument (4) that at large distances, the protoplanetary gas cannot be cooler than the cosmic background, we infer $T_G(1) = T_G(\infty) = T_{\mathrm{CB}}$, which corresponds to our ansatz (5) with $\alpha = 3$, therefore:

$$T_G(d) = T_{\mathrm{CB}} d^{-3} \quad \text{for } d < 1 \text{ au} \tag{11}$$

And:

$$T_G(d) = T_{\mathrm{CB}} \quad \text{for } \geqslant 1 \text{ au}. \tag{12}$$

Tentatively using Equation (11) with $d = R_{\mathrm{PS}} \approx 2R_S = 0.00928$ au, we deduce a very high surface temperature for the early proto-Sun: $T_{\mathrm{PS}} = 3.42\ 10^6$ K. Applying the Wien law to this temperature implies a sharp maximum of emission at wavelength 0.848 nm (1464 eV), i.e., in the soft to hard X-ray range. This is consistent with our initial hypothesis that all elements in the protoplanetary gas can be ionized according to Equations (1) and (6) and also with observation of young stars, which are emitting in the X-ray range (Montmerle et al. 2006).

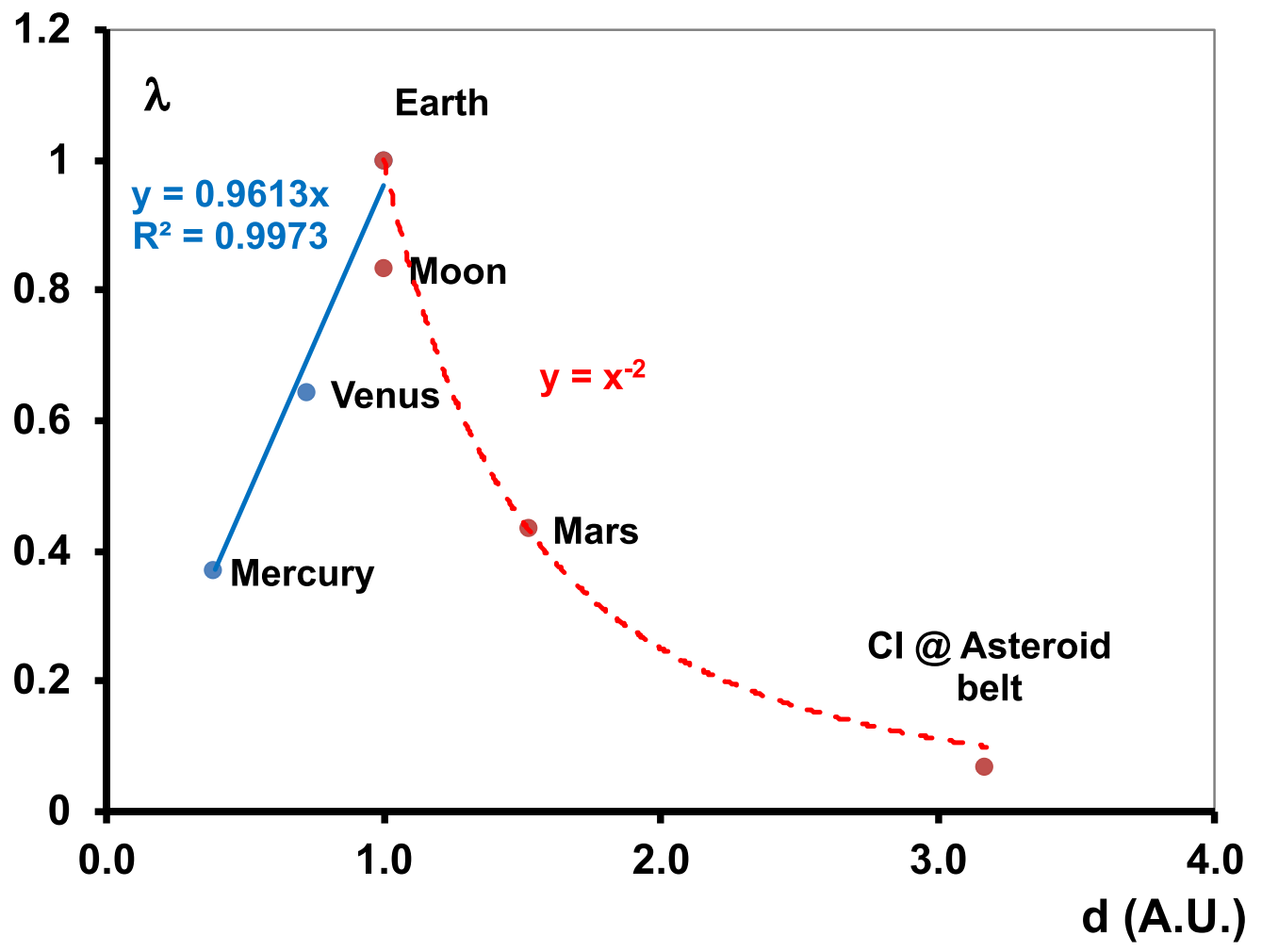

**Figure 7.** Scaling relationship observed between slopes of correlations reported in Figures 2–6 and distance $d$ to the Sun: it is linear from Mercury to Earth, as indicated by the equation and coefficient of correlation of the regression line given in the inset. The Moon, Mars, and CI asteroids lie on a $\mathbf{1/d^2}$ scaling relationship beyond Earth (broken line).

We can now consider the implications of these findings for $T_{elG}(d)$; as determined by Equation (3), it becomes:

$$T_{elG}(d) = \left(\frac{d}{R_{\mathrm{PS}}}\right)^2 \frac{1}{\pi} T_{\mathrm{CB}} d^{-3}$$
$$= \frac{T_{\mathrm{CB}}}{\pi R_{\mathrm{PS}}{}^2} d^{-1} \quad for\ d < 1 \text{ au} \tag{13}$$

$$T_{elG}(d) = \frac{T_{\mathrm{CB}}}{\pi R_{\mathrm{PS}}{}^2} d^2 \quad for\ d \geqslant 1 \text{ au}. \tag{14}$$

The electronic temperature of the nonthermal plasma is therefore minimal at $d = 1$ au, where, consistent with the previous estimates, it would reach 10,165 K. It is 91,481 K at $d = 3$ au and 109,2992 K at $d = R_{\mathrm{PS}}$. These numbers are obtained from the model while assuming $R_{\mathrm{PS}} = 2R_S$, but only the order of magnitude should be taken into account since solely the order of magnitude of the actual value of $R_{\mathrm{PS}}$ can be guessed. The corresponding electron speeds imply negligible relativistic effects with Lorentz factors $\gamma = (1 - (v_e/C)^2)^{-1/2}$ exceeding 1 by less than $2 \times 10^{-4}$. Estimating the average density of electrons in the protoplanetary disk at $8.2 \times 10^{19}$ m$^{-3}$, the Debye length of the plasma would remain of the order of a few microns, corresponding to a quasi-neutral nonthermal plasma.

We are entitled to compare the functional dependency of Equation (13) with that of the kinetic energy of an orbiting electron in gravitational equilibrium at distance $d$, which is equal to minus its gravitational potential energy:

$$k_B T_{elEq}(d) = G M_{\mathrm{PS}} m_e^0 \gamma d^{-1}. \tag{15}$$

Where $G$ is the gravitational constant, $M_{\mathrm{PS}}$ the proto-Sun mass, $m_e^0$ the mass of an electron at rest, and $\gamma$ the Lorentz factor. Identifying $T_{elEq}(d)$ of Equation (15) and $T_{elG}(d)$ of

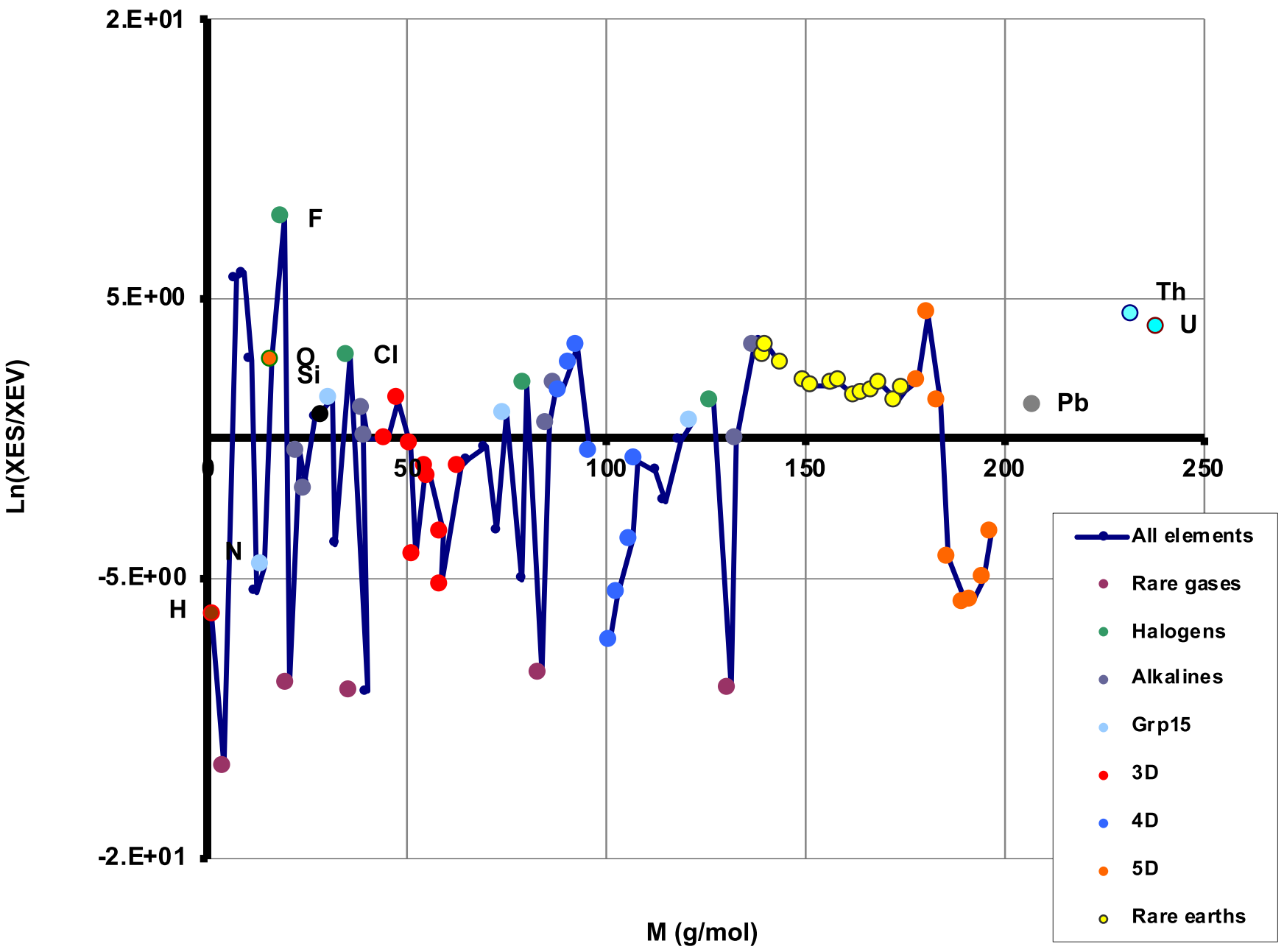

**Figure 8.** Plot of partition coefficients between Earth surface and Earth volume as a function of atomic mass. Numbers in ordinates (listed in Table A2) are computed from Equation (16), so that they essentially compare the experimentally observed crustal (or surface) relative abundances and the predicted initial bulk relative abundances on the basis of Equation (6) with $R_{PS} \approx 1.5R_S$ ensuring almost constant partition coefficients for noble gases. Elements with positive partition coefficients are enriched at the surface with respect to the bulk, while elements with negative partition coefficients are enriched in the bulk with respect to surface.

Equation (13) leads to an estimate of $M_{PS} \sim 0.173\, M_S$. Together with the above estimate of its surface $T_G$ of order $10^6$ K, this is consistent with the proto-Sun following the Hayashi track on its rapid evolution toward the onset of nuclear fusion, giving birth to the contemporary Sun. We can then infer that for $d < 1$ au, electrons excited in the plasma are close to gravitational equilibrium, while they have an excess of kinetic energy for $d \geqslant 1$ au.

## 4. Radial Differentiation inside Earth and Planets: Chemistry versus Gravity

Let us now examine the origin of outliers in the correlations presented in Figure 1 and in Appendix Figures A6–A10. The occurrence of these outliers significantly affects the quality of correlations and therefore seemingly weakens our model. We will show that it is not the case, and moreover, the departures from the law expressed by Equation (6) actually convey crucial information about the partition of elements between the surface and the inner materials of planets.

In Figures 1 and A6–A10, differentiation factors of the elements are calculated after experimental elemental abundances measured and averaged for the crusts, and eventually the hydrosphere and atmosphere, i.e., planet surfaces, while Equation (6) should hold for the relative abundances averaged over the bulk planetary-volume relative abundances. Accordingly, departures from the law described by Equation (6) (vertical distances in log scale to the regression line) straightforwardly convey the information on the ratio of surface to volume concentrations. This is expressed by Equation (16) demonstrated in the Appendix:

$$\ln\left[\frac{X_{\mathrm{ES}}}{X_{\mathrm{EV}}}(M)\right] = [\ln f_{\mathrm{ES}}(M) - \ln f_{\mathrm{EV}}(M)] + \ln\left[\sum_M X_{\mathrm{ES}}(M)\frac{f_{\mathrm{EV}}(M)}{f_{\mathrm{ES}}(M)}\right] \tag{16}$$

where subscripts ES and EV here stand for Earth surface and Earth volume, respectively, but are valid for any planet. Hence, crust data points located above the regression line stand for elements that are enriched at the surface relative to volume (e.g., F, O, Si, P, B, Cl...), or added a posteriori from outer space to surface, whereas those located below stand for elements that are depleted in surface, relative to volume (e.g., Fe, Co, Ni, Cr, H, ...) or might also be lost through escape to space by the Jeans effect (notably H, He).

In the case of Earth, plotting those partition coefficients $\ln\left(\frac{X_{\mathrm{ES}}}{X_{\mathrm{EV}}}(M)\right)$ (provided in Table A2 of SI), against atomic mass (Figure 8), reveals distinctive periodic trends in the depletion or enrichment in chemical elements at the Earth surface with respect to Earth volume:

1. Rare earth elements are systematically more abundant at the surface, with similar partition coefficients. (This is in full agreement with known incompatibility diagrams established for crust rocks and OIB basalts with respect to MORB-N basalts).
2. The most electronegative elements, O and halogens, are much more abundant at the surface than in volume. Halogens are the sole elements with positive heats of formation of their oxides, while it is zero by convention for $O_2$ (g) itself.
3. Transition elements in the 3d, 4d, and 5d series exhibit complex periodic patterns, with the elements from groups

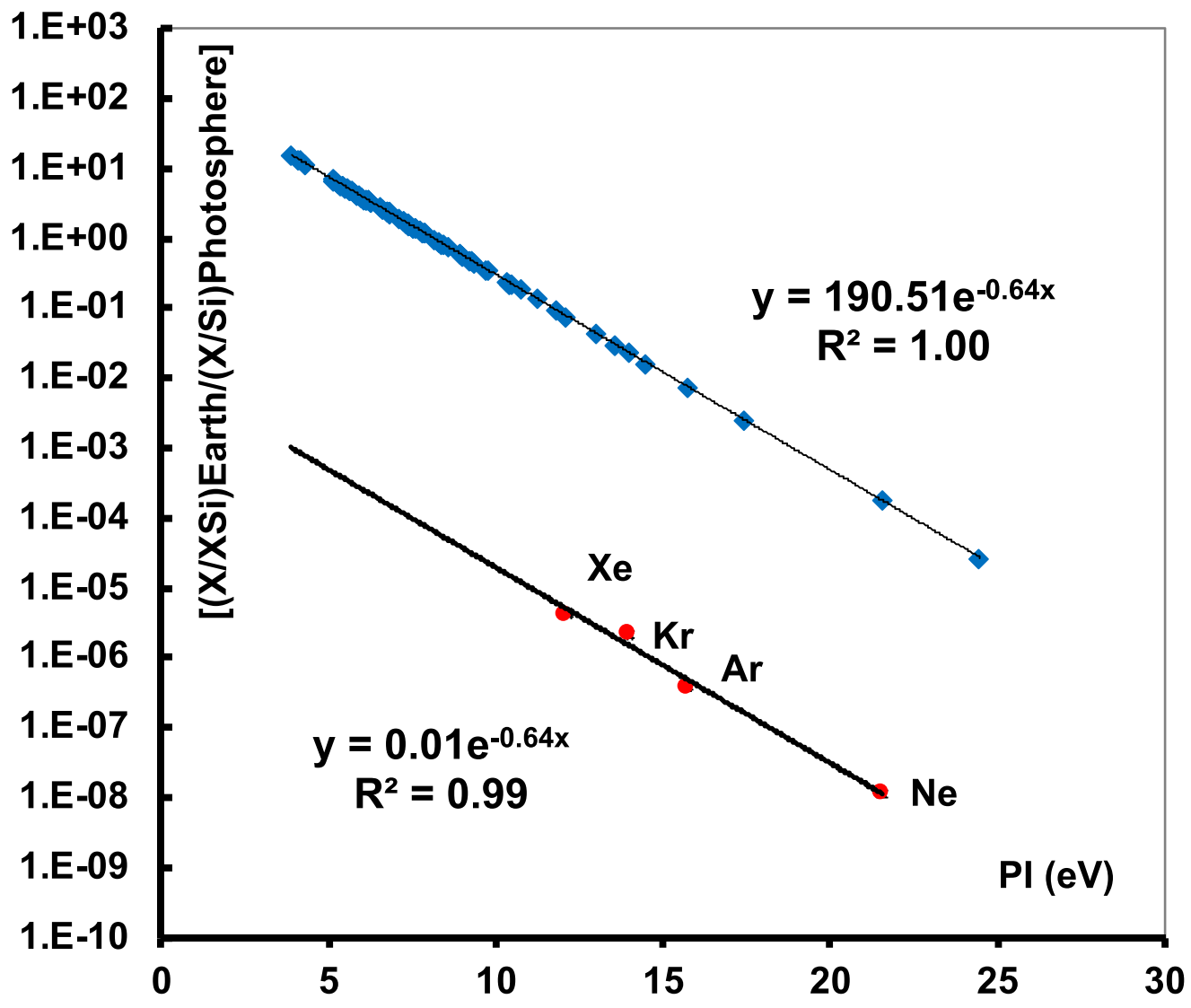

**Figure 9.** Plot of differentiation factors for noble gases vs. PI. Blue diamonds account for Equation (6) with $R_{PS} \approx 1.5R_S$ and all partition coefficients equal to one. This reference line has a slope that equalizes partition coefficients of noble gases.

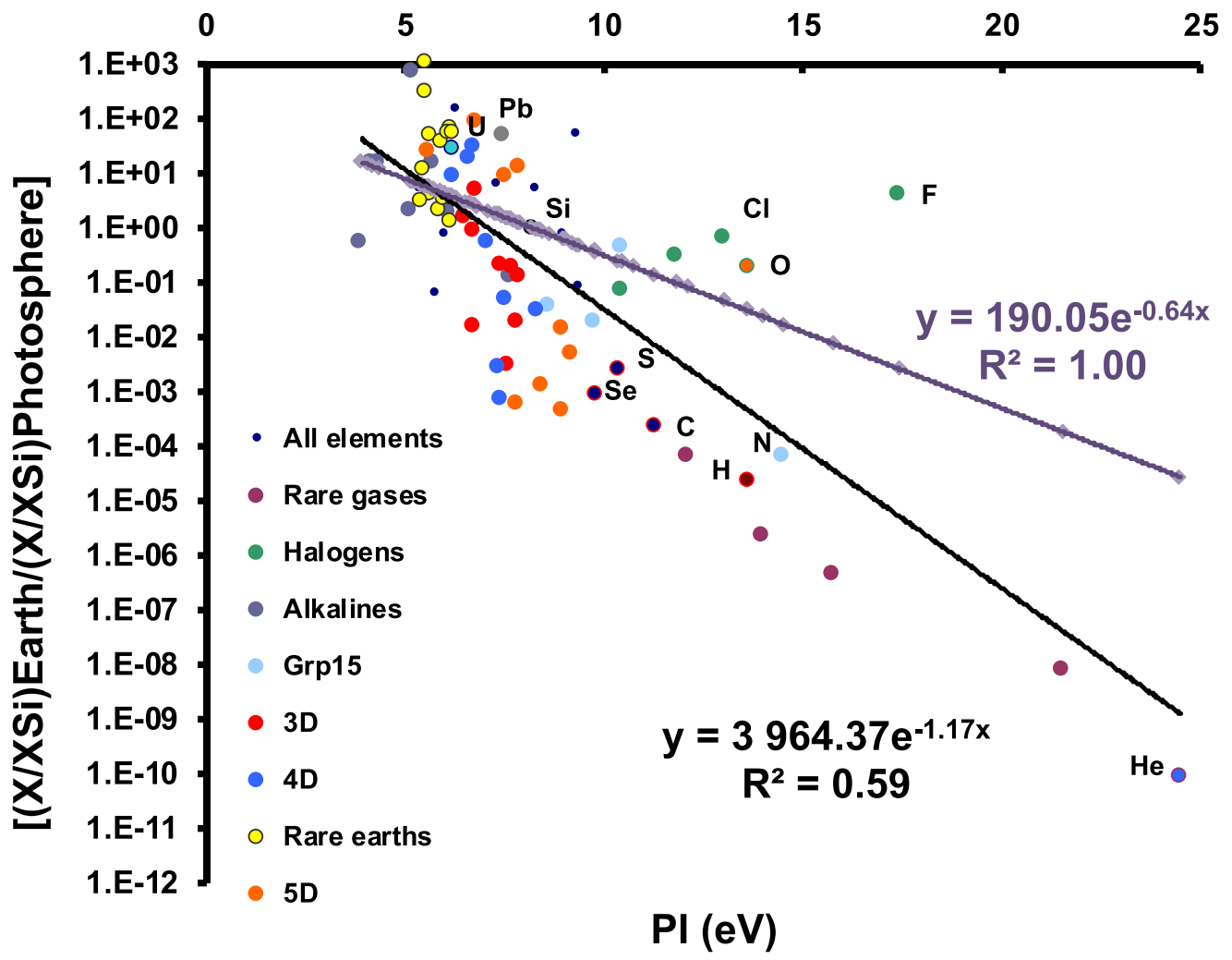

**Figure 10.** Same as Figure 1, with the plot of Equation (6) (magenta diamonds and equation of regression lines in inset) superimposed, assuming equal partition coefficients for noble gases, which in turn determines $R_{PS} \approx 1.5R_S$.

3 to 5 enriched at the surface, then increasing depletion in favor of volume with d-band filling, going through a maximum for Ni, Ru, and Os, then decreasing again.

Figure 9 has the same axes as Figure 1, but is restrained to noble gases, except He: differentiation factors for the latter are shown linearly correlated to their ionization potentials. Since their partition coefficients should be independent of chemical interactions, it can be inferred that they provide the reference slope for Equation (6), which, in other words, determines the actual value of the proto-Sun radius $R_{PS} \approx 1.5R_S$. Figure 10 allows comparison of the resulting plot of Equation (6) assuming all partition coefficients equal to one, to the regression line of Figure 1 (note that Figure 8 was built with $R_{PS} \approx 1.5R_S$). Accordingly, the actual proto-Sun temperature would be $T_{PS} = 8.09 \times 10^6$ K. Applying the Wien law to this temperature implies now a maximum of emission at wavelength 0.358 nm, i.e., in the slightly harder X-ray range than our first guess. The minimal nonthermal plasma electronic temperature at $d = 1$ au is now 18,070 K, and the proto-Sun mass satisfying Equation (15) is $M_{PS} \sim 0.309M_s$.

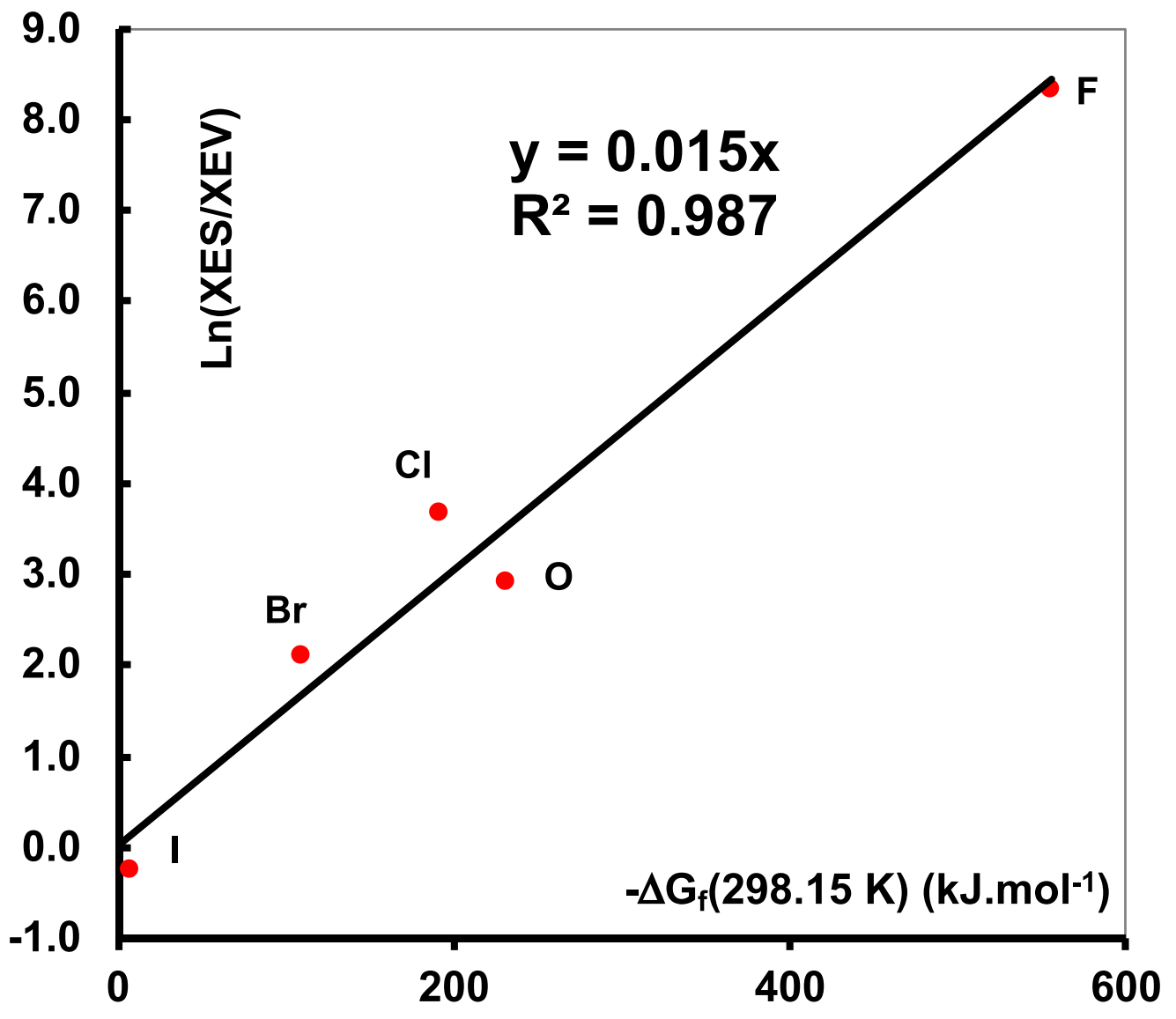

**Figure 11.** Correlation of the present time partition coefficients of oxygen and halogens with their enthalpies of hydruration. This plot suggests that these most electronegative elements were chemically driven to the Earth's surface by a flux of hydrogen.

With all parameters in Equation (6) determined, we note that our theory assigns a very peculiar position to our planet Earth: with the maximum $\lambda(d = 1 \text{ au}) = 1$, the Earth is the most chemically differentiated planet in the solar system, exhibiting the most negative slope in diagrams such as Figures 1 or 10. The corresponding diagrams for chondrites considered to be samples from the asteroïd belt, and for the other inner planets of our solar system, are presented in Figures A6–A10 in the Appendix. In these figures, the available experimental differentiation factors (representative of planes surfaces) are plotted as red dots, and the theoretical law as a black line (representative of the bulk planet in average). Above these lines lie elements segregated in surface, and below are elements depleted from the surface, showing trends of partition coefficients similar to those of Earth. In view of the much less statistically significant experimental differentiation factors available for planets other than Earth, we did not attempt to interpret further the partition coefficients that can be deduced.

We present further interpretations of trends for partition coefficients for the Earth in terms of thermochemical properties of the elements. First, considering oxygen and halogens, Figure 11 shows the striking linear correlation of the observed partition coefficients between the crust and bulk minus the enthalpies of formations of their hydrides, i.e., $H_2O$, HF, HCl, HBr, and HI.

This almost perfect correlation allows us to propose that these mostly electronegative elements were chemically driven to the Earth's surface by a flux of hydrogen along geological times. This finding also suggests an explanation for the so far mysterious fact that oceans are so salty (we could find no scientific publication quantitatively addressing this question) since they would result from a progressive upsurge of acidic water prone to be neutralized by reaction with surficial, basic alkaline rocks. Hydrogen-driven oxygen enrichment of the

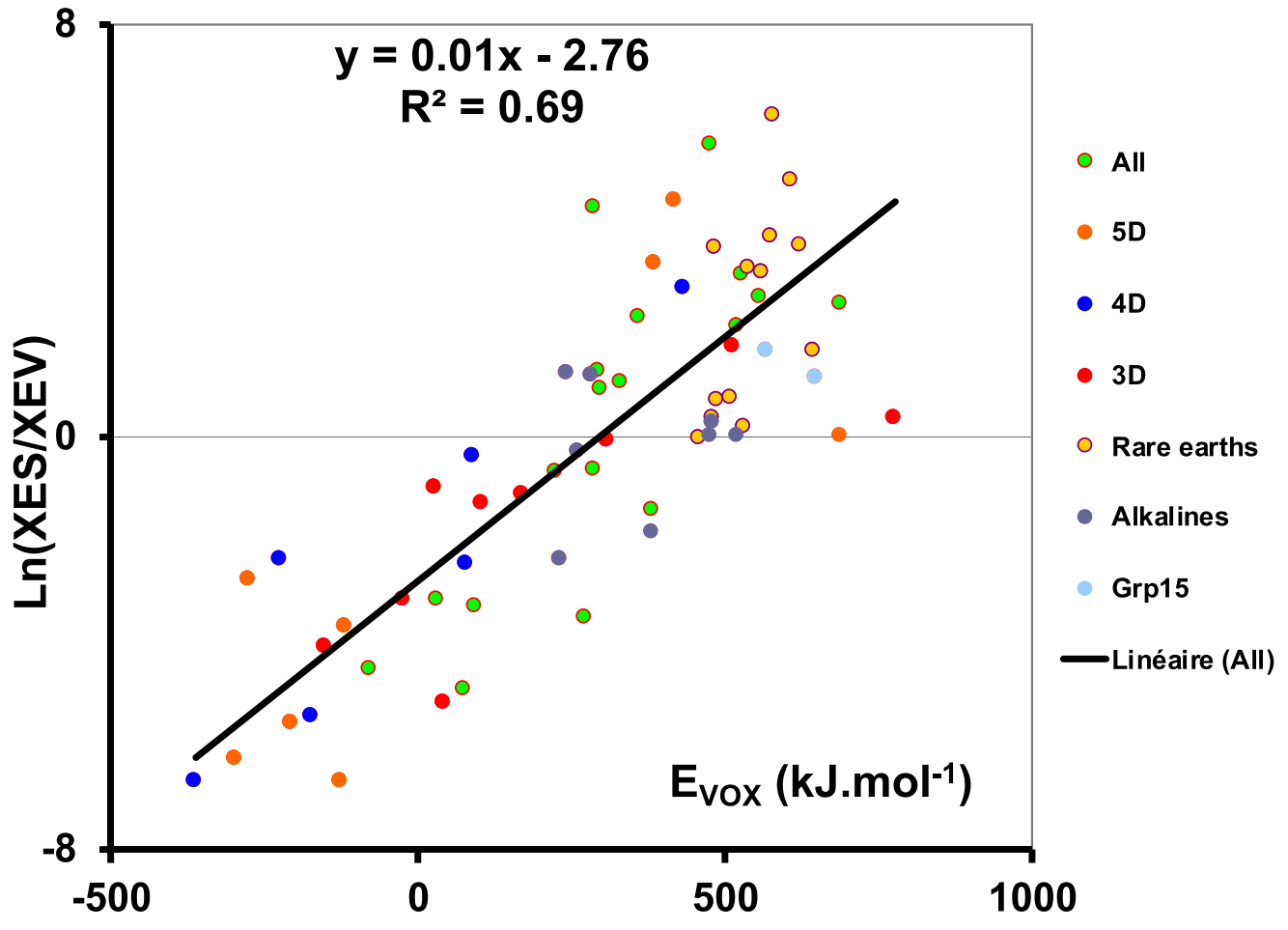

**Figure 12.** Correlation of the present time partition coefficients of elements other than oxygen and the halogens, minus their enthalpy of oxidation at 4100 K.

Earth's surface was at the origin of the further segregation of some elements based on their affinity to oxygen, which is traditionally explained as the "siderophile" versus "lithophile" element differentiation.

Figure 12 shows the fair correlation obtained plotting partition coefficients of all the other elements (up to U) against the enthalpy of formation of their oxide at a very high temperature, 4300 K, which corresponds to a maximal coefficient of correlation. Therefore, chemical elements are segregated radially from inner Earth toward its crust according to a measure of their affinity with oxygen, following the gradient pre-established by the outward hydrogen flux. The latter is expected as a consequence of the Jeans effect, i.e., the fact that dihydrogen gas is unstable with respect to Earth's surface gravity at the average Earth surface temperature and tends to escape to space. On its way to the surface through the mantle, the flux of hydrogen is "purifying" it from the electronegative elements, a mechanism similar to those used in hydrogen metallurgy, like smelting reduction and metal purification by hydrogen.

This correlation is also consistent with the fact that less common transition metal ore deposits are linked to ancient or still active volcanic systems (Gibson 2005), i.e., conveyed from inner Earth by magmatic upsurges. Lastly, elemental carbon is predicted to be more abundant in the inner planet than its surface, and indeed, diamonds are generally included in kimberlites, which are of deep origin. This again raises the controversial question of an abiogenic source of at least a part of the hydrocarbons disseminated in the crust under various forms and H/C ratios.

## 5. Discussion and Implications

According to Equations (6) and (16), considering the elements heavier than Ni as traces, and making use of available experimental relative abundances for the Earth's crust, initial bulk mass fractions of matter for proto-Earth can be calculated (Table A1 in the Appendix, column (a)). This calculation predicts a very high initial content of H, major element ranking as:

$$\mathrm{H} > \mathrm{Fe} > \mathrm{Mg} > \mathrm{Si} > \mathrm{C} > \mathrm{Al}$$
$$> \mathrm{Na} > \mathrm{Ca} > \mathrm{O} > \mathrm{Ni} > \mathrm{S} > \mathrm{K} > \mathrm{Cr}.$$

The bulk mass fractions we deduce reveal that the proto-Earth has been more H-rich and O-poor than currently considered. In quantitative terms, Equation (6) yields up to an initial hydrogen content of 83.3 wt% for the bulk material of the proto-Earth. However, most of it most likely escaped during the accretion stage, except for a fraction of H that was chemically and physically bonded.

Indeed, multiple studies suggest that the Earth's interior has been, and could still be, rich in hydrogen (see, for example, the recent review Zgonnik 2020). Yet, iron hydride is more considered as a key ingredient of Earth's core (Badding et al. 1991; Williams & Hemley 2001; Isaev et al. 2007; Rumyantsev 2016) to resolve the so-called "core density-deficit problem." For example, Murphy suggests that an $\mathrm{FeH}_{0.14}$ composition for the inner core produces a reasonable agreement with seismic observations (Murphy 2016). Another study showed that an alloy $\mathrm{Fe}_{0.88}$ $\mathrm{Si}_{0.12}$ $\mathrm{H}_{0.17}$ containing 0.32 wt% hydrogen was reported to match density and seismic profiles for the outer core (Tagawa et al. 2016). A very recent study proposed that the core might contain 80 times more hydrogen than the mass of all hydrogen in the ocean (Ikuta et al. 2019). However, the idea of a hydrogen-rich core remained marginal (despite a good fit between experimental results with iron hydrides and theoretical studies) because of the lack of a plausible mechanism of delivery of large quantities of hydrogen into proto-Earth. Our current study covers this gap.

Assuming H has been retained until present day as hydrides of transition elements in the deep mantle and core conditions, stoichiometries allow us to estimate a content of the order of 4 wt% combined H, i.e., $\sim 1.2 \times 10^{11}$ Gt $\mathrm{H}_2$ (Table A1 in the Appendix, column (c)), which is in the same range as the previous estimate (Larin 1993). This huge amount, if available as fuel, would represent in ~6 Gyr of the current primary energy consumption worldwide (available atmospheric oxygen would then become limiting).

Hydrogen stored in the Earth as hydrides could progressively decompose and sustain the continuous degassing of hydrogen over geologic time, especially on the early stages of the planet's history (Larin 1993; Walshe et al. 2005; Gilat & Vol 2012). The relatively low escape velocity of molecular hydrogen at Earth's surface favors its diffusion toward outer space in the absence of oxidizing agents in the atmosphere (as suggested for early Earth). Indeed, it was proposed that an Archean atmosphere could have contained up to 30% hydrogen (Wordsworth & Pierrehumbert 2013) and that hydrogen losses on early Earth were six orders of magnitude greater than present day loss estimates (Pepin 1991).

Even assuming all H and He have escaped along geological times (Table A1 in the Appendix, column (b)), the ranking of the remaining elements is unchanged, but their mass fractions are of course much higher (Fe ~ 30 wt%, O ~ 4 wt%). A predicted global O content for Earth of ~4 wt% seems difficult to accept since O is obviously the most abundant element on Earth's surface (see Table A2 in the Appendix). But one must realize that there is no direct proof that it is so for the Earth in average (that oxygen is abundant in the Earth's interior from where we do not have any statistically representative samples). Remember that Earth's lithosphere accounts for merely about 0.5% of the total mass of Earth and 1.6% of its radius. Further, one should keep in mind that at the latest stage of accretion, terrestrial planets were impacted at high kinetic energies by silicate-rich large bodies formed in farther regions of the solar system. These cruisers were engulfed across the relatively young crust and their material directly incorporated to inner Earth. This might have significantly enriched the actual inner Earth in O and other elements, while our predictions would then hold for an earlier stage of the planets' differentiation.

Including this idea and the prevalence of hydrides in the deep inner planet, our next report in preparation will present a mineralogical model of inner Earth making use of extensive first principles (density functional theory) calculations of mineral densities and elastic properties in order to match our predictions with geophysical interpretations of seismic data such as the Primary Reference Earth Model.

We have shown the predicted strong surface enrichment for lanthanides, U and Th (which allows us to assume that it is also the case from all actinides). It has a strong consequence on the contribution of long-period radioactive isotopes of actinides to the heat flux generated inside Earth, which should be minimized accordingly. Our results give way to such calculations, but we will leave it as out of scope of the present report. On the contrary, our model predicts ∼1 wt% K, a much higher content of inner Earth in K (among other monovalent alkalines) than usually assumed. Therefore, the radioactive decay of the inner amount of $^{40}$K (presently 117 ppm moles in K (Bohlke et al. 2005)) should provide most of the thermal energy generated inside the Earth, i.e., a heat flux of ∼200 TW. It might also open questions concerning mechanisms by which such an amount of heat is dissipated (Bezrukov et al. 2018; Barabanov et al. 2019). Also, radioactive decay of $^{40}$K, $^{238}$U and $^{232}$Th generate a flux of geo-neutrinos. A recent study shows that the experimental data from the Borexino does not contradict our model, which proposes significantly higher content of those elements for the bulk Earth (Bezrukov & Sinev 2015; Sinev et al. 2015).

The present results suggest that Earth's mantle and core can potentially be more reduced than usually considered and hydrogen is more evenly distributed in the solar system than previously thought. A very recent review (Zgonnik 2020) conclusively shows that hydrogen is much more widespread on Earth than previously thought and that a deep-seated hydrogen is likely the largest source of natural hydrogen. Many studies hypothesize that Earth's interior is rich in hydrogen to explain various natural phenomena: volcanism and formation of mineral systems (Walshe et al. 2005; Gilat & Vol 2012), the chemistry of the atmosphere and oceans and climate change (Syvorotkin 2010), and much more (Zgonnik 2020).

In conclusion, we propose that a magnetically driven chemical gradient at the scale of the Solar nebula has imprinted chemical differentiation of planets in the further accretion stages and propose a quantitative model for it. The model provides an opportunity to address in a new light many topics with issues: differences in compositions of rocky planets, characteristics of the proto-Sun and scaling law for temperature inside the protoplanetary nebula, the initial composition of planetary materials, planets and chondrites in the solar system, the driving forces for radial differentiation, the origin of volatiles, etc. The model also provides a means to estimate the chemical composition of exoplanets based on the spectral data of their stars. Finally, our model predicts that the Earth's elemental composition is drastically different from current estimations (especially for hydrogen), with many consequences beyond the scope of this paper, notably a potential future supply of clean primary energy.

H.T. developed the equations, performed the tests against experimental data, and wrote the first drafts. V.Z. gathered literature data and contributed to numerical tests. Both authors planned the research, contributed to critiquing and improving the interpretation, and to writing the final form of the manuscript.

We warmly thank Valérie Beaumont for her contribution. This research did not receive any specific grants from funding agencies in the public, commercial, or not-for-profit sectors.

# Appendix A
# Supplementary Methods

## *A.1. Calculation of the Differentiation Factor for Earth*

In addition to average crust, the relative abundances were corrected using data from the hydrosphere and the atmosphere

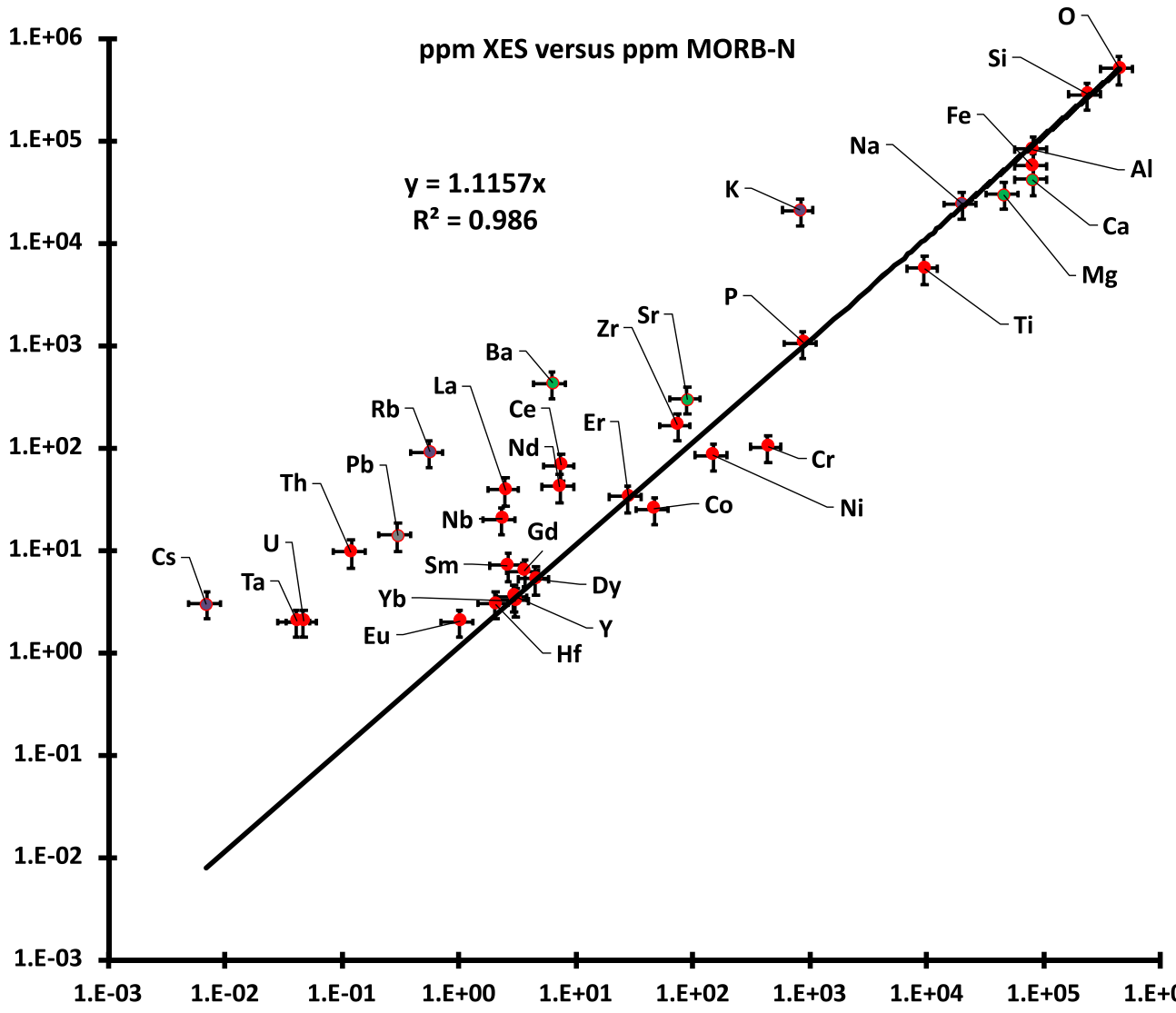

**Figure A1.** Earth surface concentrations, in ppm, based on continental crust data from Lide (2005) plotted against average concentrations measured in samples of oceanic crust rocks MORB-N (middle oceanic-ridge basalt type-N) from Sun and McDonough (1989).

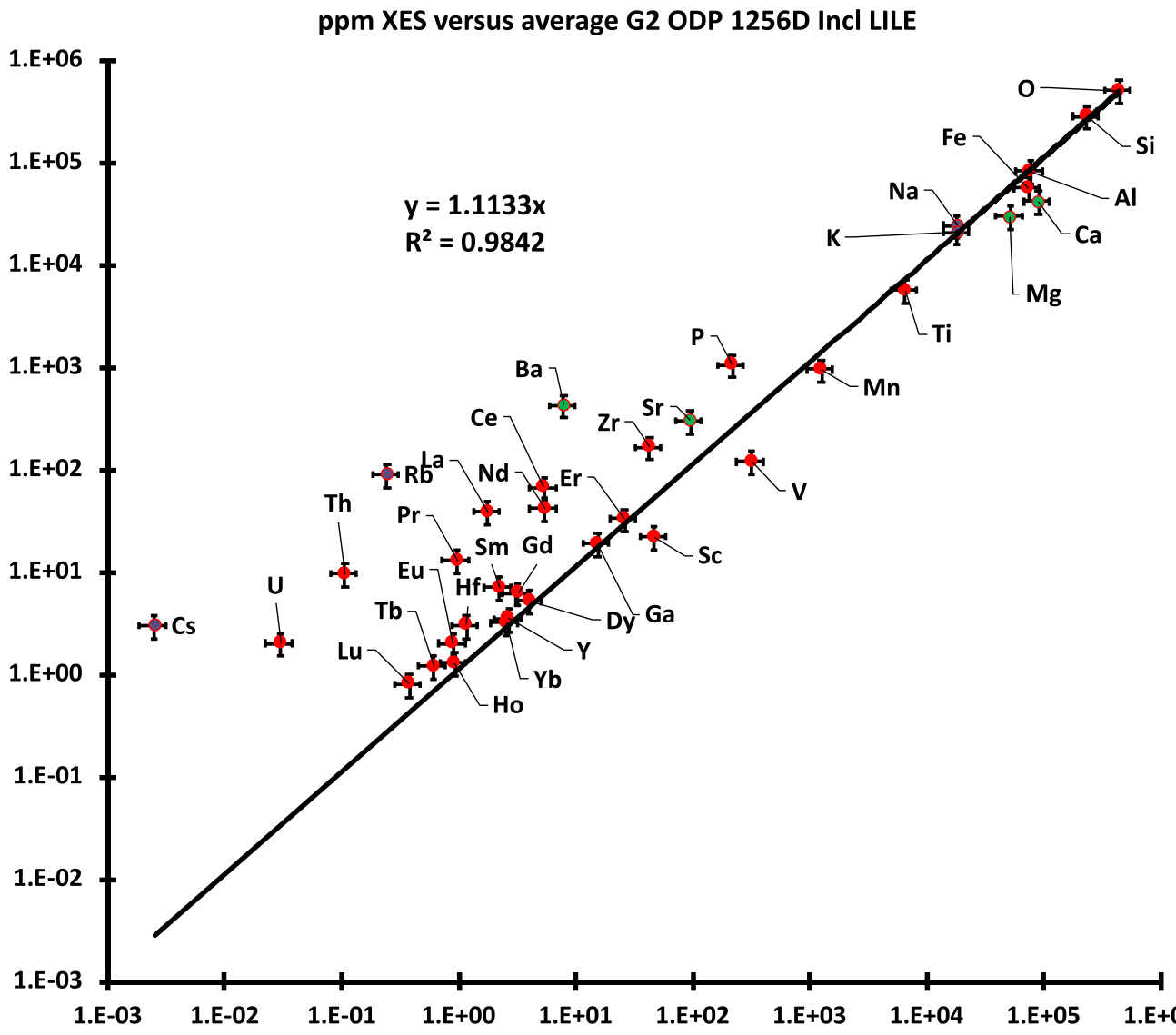

**Figure A2.** Earth surface concentrations, in ppm, based on continental crust data from Lide (2005) plotted against average concentrations measured in samples of oceanic crust rocks G2 from the east Pacific including LILE (large ions lithophile elements; Yamazaki et al. 2009).

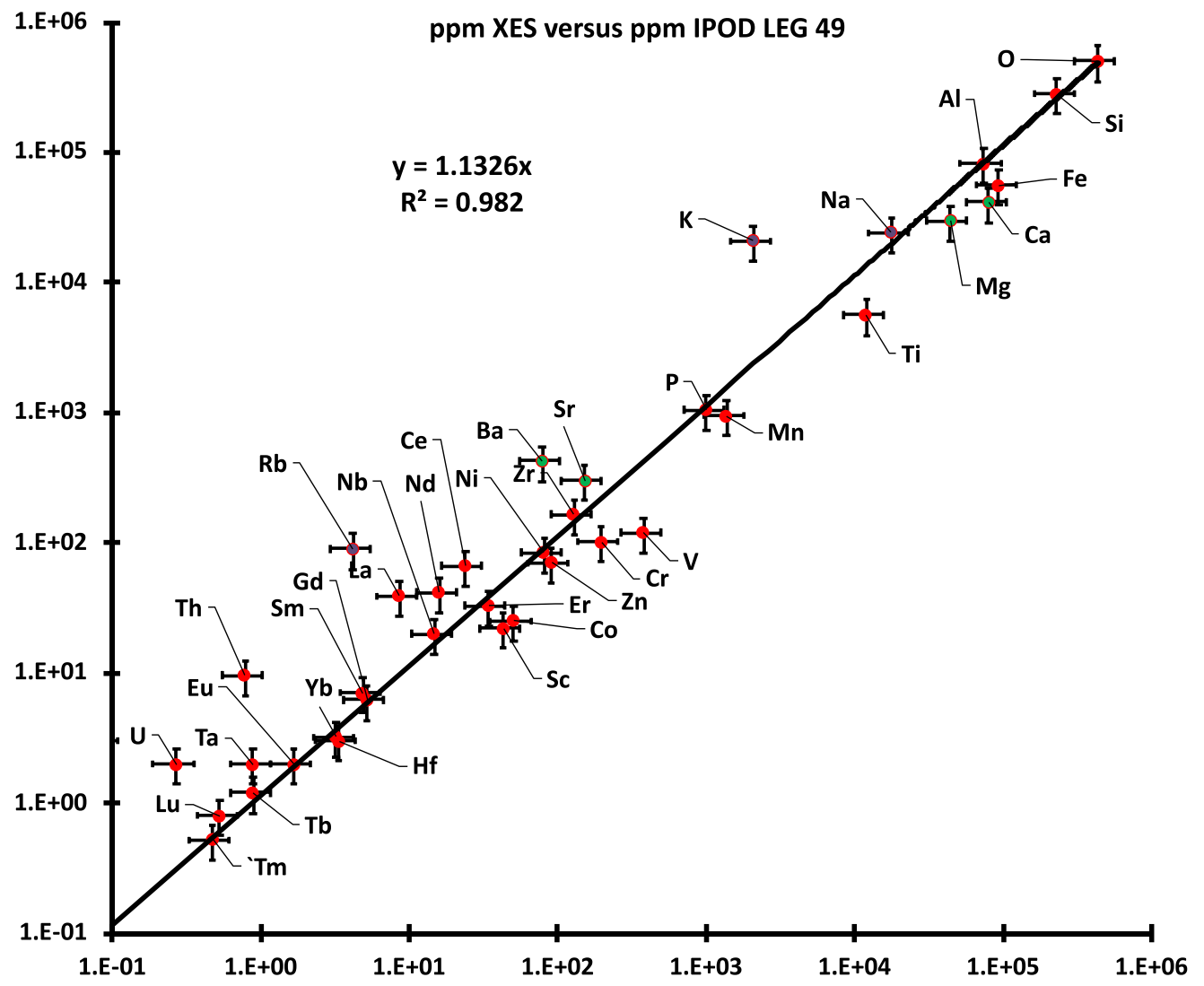

**Figure A3.** Earth surface concentrations, in ppm, based on continental crust data from Lide (2005) plotted against average concentrations measured in samples IPOD LEG 49 of oceanic crust rocks from the north Atlantic (Wood et al. 1979).

(see the CRC Handbook reference, Lide 2005, in the main text). Special attention is given to radiogenic nuclides. Helium is not considered in calculations as, in addition to primordial $^{3}He$ and $^{4}He$, $^{3}He$ is produced from $^{6}Li$ while $^{4}He$ is produced during decay of different radionuclides. In these conditions, significant differentiation factors cannot be calculated for He. The Earth value for argon abundance uses $^{36}Ar$ but excludes $^{40}Ar$ which is essentially radiogenic and produced from $^{40}K$, with Ar and K displaying very different IPs (15.75 and 4.34 eV, respectively). Xe abundances are not corrected from radiogenic $^{129}Xe$ produced from $^{129}I$, since Xe and I display similar IPs (12.13 and 10.45 eV, respectively). The contribution of radiogenic $^{21}Ne$ is considered negligible and therefore Ne abundances were not corrected. Li and Be that are burnt during nucleosynthetic reactions in the Sun are not reported (B might have been also affected).

### A.2. Comparing Continental versus Oceanic Crust Data

In Figures A1–A3 below, we have plotted in ordinates the Earth surface concentrations, in ppm, we used for our work, based on continental crust data from Lide (2005), against the measured concentrations of major, minor, and trace elements, in ppm, in abscissae, for actual samples of oceanic crust rocks (respectively, MORB-N for middle oceanic-ridge basalt type-N; Sun and McDonough (1989), G2 from the east Pacific (Yamazaki et al. 2009), and IPOD LEG 49 from the north Atlantic (Wood et al. 1979); references are provided herein). In all cases, we obtain excellent linear correlations with squared coefficients of correlation above 0.98 slopes 1.11–1.13. Outliers are LILE (large ion lithophile elements), i.e., with the lowest ratios charge over ionic radius $q/Ri$ (below 0.007 for K+, Rb+, Cs+, and of the order 0.009 for the heaviest rare earths at the 3+ oxidation state), which as is well known, tend to be excluded from octahedral and tetrahedral sites of silicates and in general oxides. At the same time, the major elements that are dominating the correlation fit perfectly on the line.

These excellent correlations mean that the mass fractions normalized to Si for these actual oceanic crust samples are indistinguishable from those used to build Figure 1 of the article. Therefore, we can conclude that model-independent estimations of crustal relative abundances of elements are indeed reliable, even if based on continental crust data.

The Figures A1–A3 include 30% error bars in abscissae: these do not reflect analytical instrumental uncertainties, but rather the dispersion between various samples of the same family of rocks and various laboratories for the same samples.

### A.3. Correlation of Differentiation Factors to Condensation Temperatures of the Most Stable Minerals

In the following Figure A4 we present the existing correlation between first ionization potential (FIP) and volatility of most stable mineral phases of elements likely to condense in the protoplanetary nebula (expressed as midpoint condensation temperatures, as reported by Lodders (2003), while in Figure A5 we plot the correlation between this measure of volatility versus relative abundances. Figure A4 shows that although volatilities and FIP are poorly correlated, with $R^2 = 0.458$ (actually, volatility is rather determined by the cohesive energy of the element in its condensed state, which, in general, has little to do with the FIP since bulk solid compounds of elements are not generally comparable to a network of N monovalent ions sharing N delocalized valence electrons). Figure A5 shows a poor correlation, with $R^2 = 0.40$, including noble gases and $R^2 = 0.20$ otherwise. This has to be compared with Figure 1 in the main text, where relative abundances and FIP are fairly correlated with $R^2 = 0.59$ including noble gases up to He. We conclude that our proposal does improve the ordering of data.

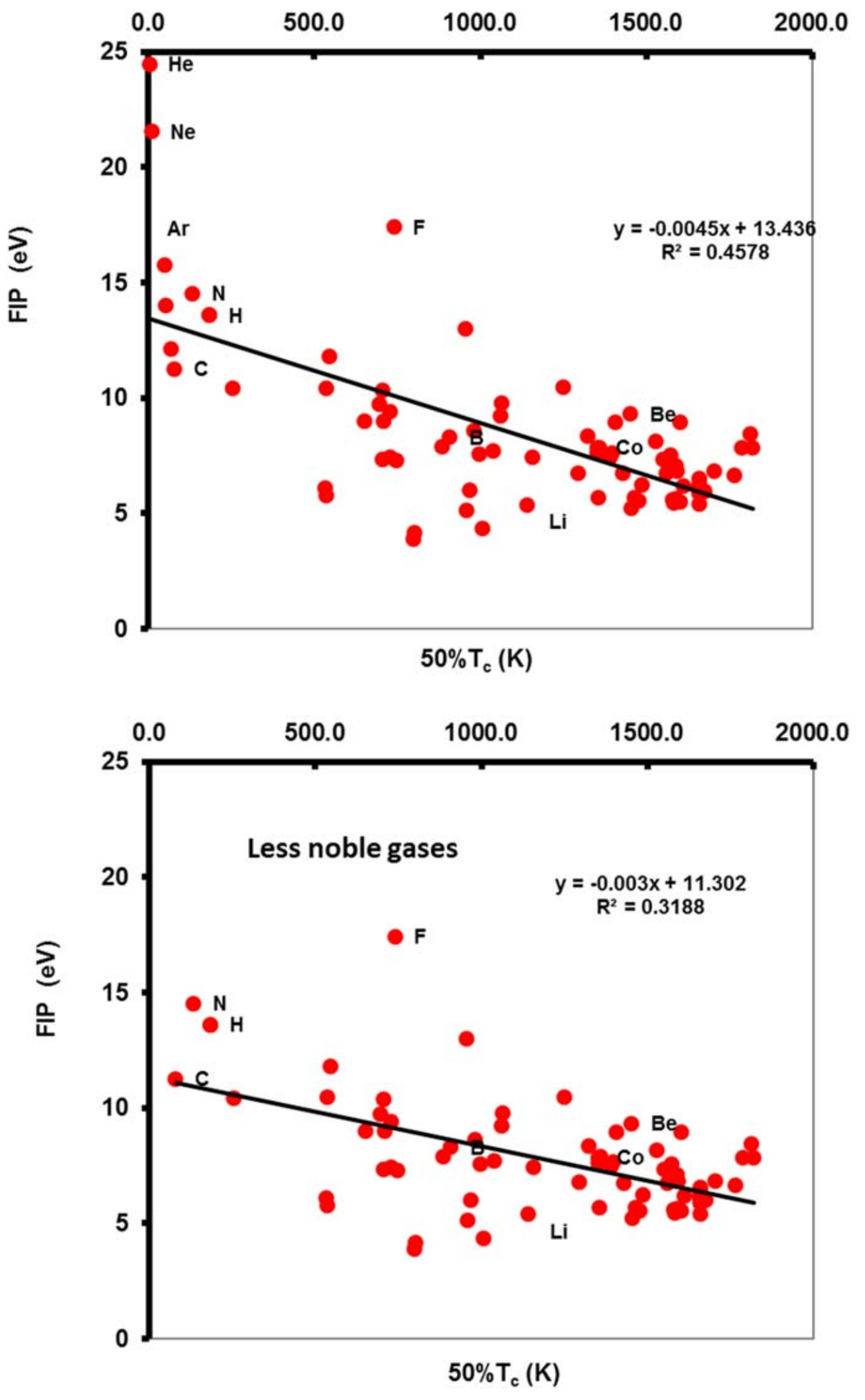

**Figure A4.** Correlation between first ionization potential of elements (H to U) and temperature of condensation of 50% of the mass into the most stable mineral, 50% $Tc$, as reported by Lodders (2003). Top—with noble gases; bottom—excluding noble gases.

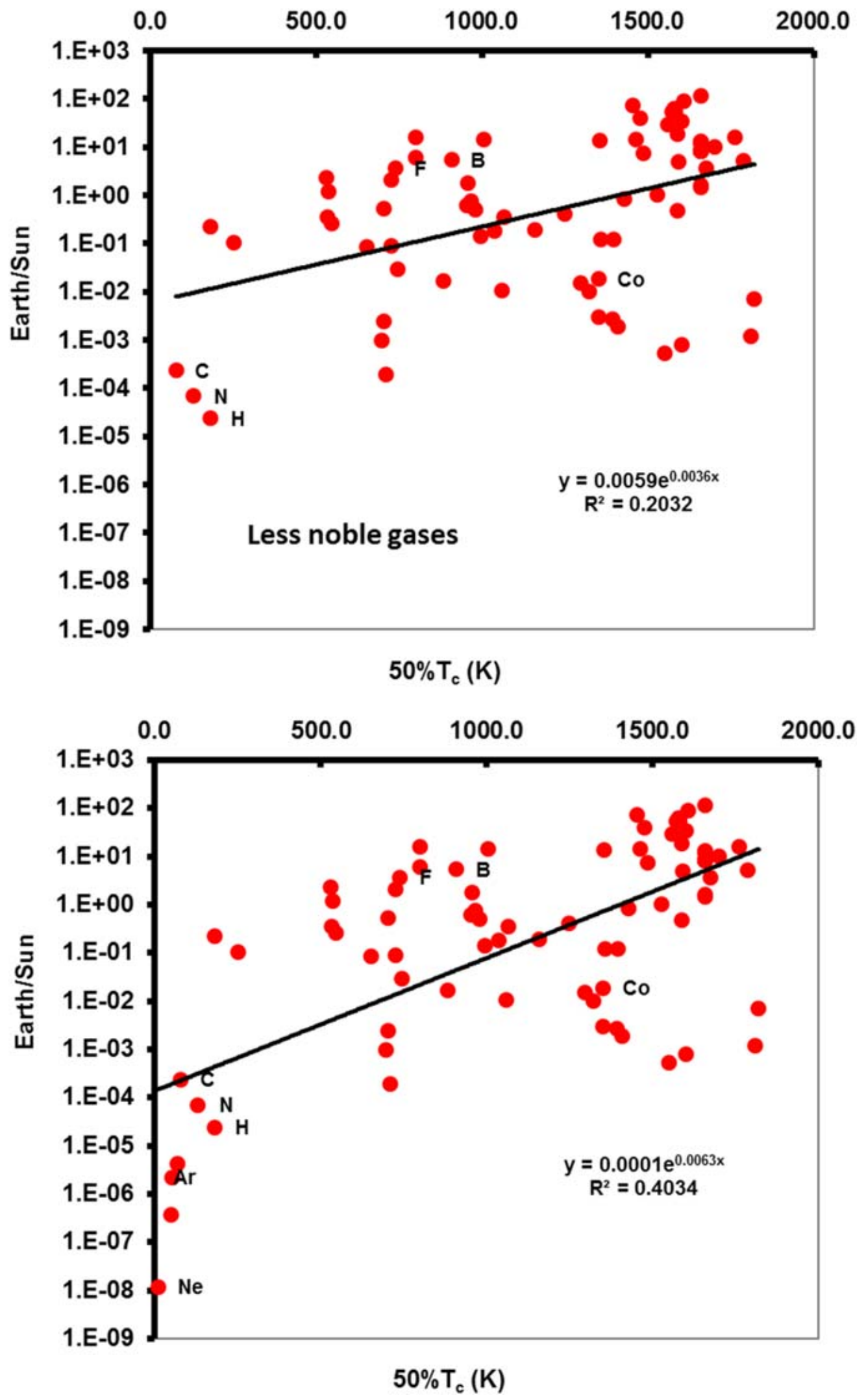

**Figure A5.** Correlation between relative abundances of elements (H to U) and temperature of condensation of 50% of the mass, 50% *Tc*, as reported by Lodders (2003). Top—excluding noble gases; bottom—including noble gases.

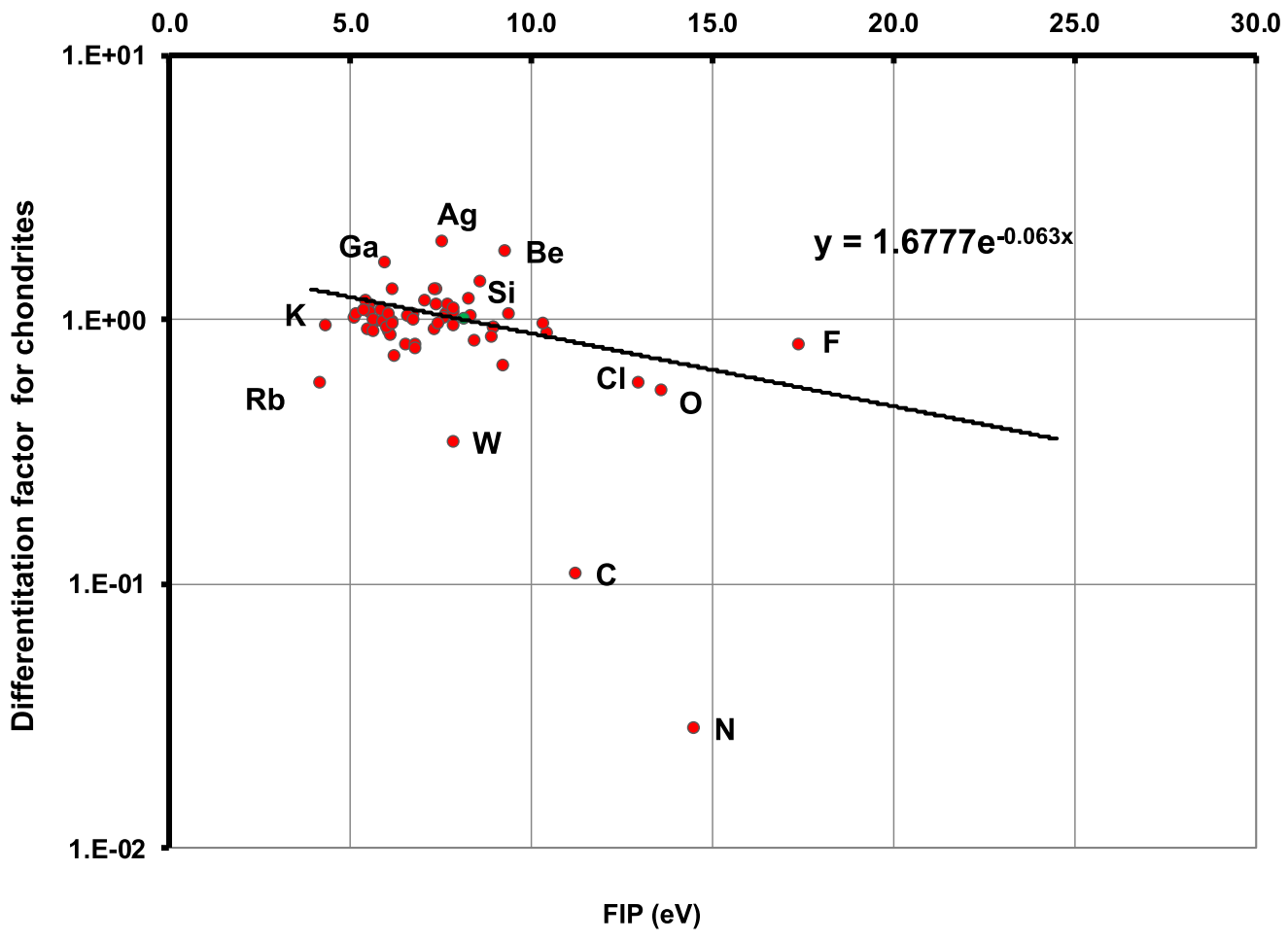

**Figure A6.** Correlation between the observed differentiation factors for chondrites assumed representative of the asteroids belt, and first ionization potential (FIP) in a semilogarithmic plot. The black line is the theoretical differentiation factor given by Equation (6) with $R_{PS} \approx 1.5 R_S$ $d = 3.17$ au and $T_G(d) = T_{CB}$ (its equation in inset). Elements C and N might have been in part burnt off when the meteorites entered the atmosphere of Earth, while additional O was incorporated into the bulk material.

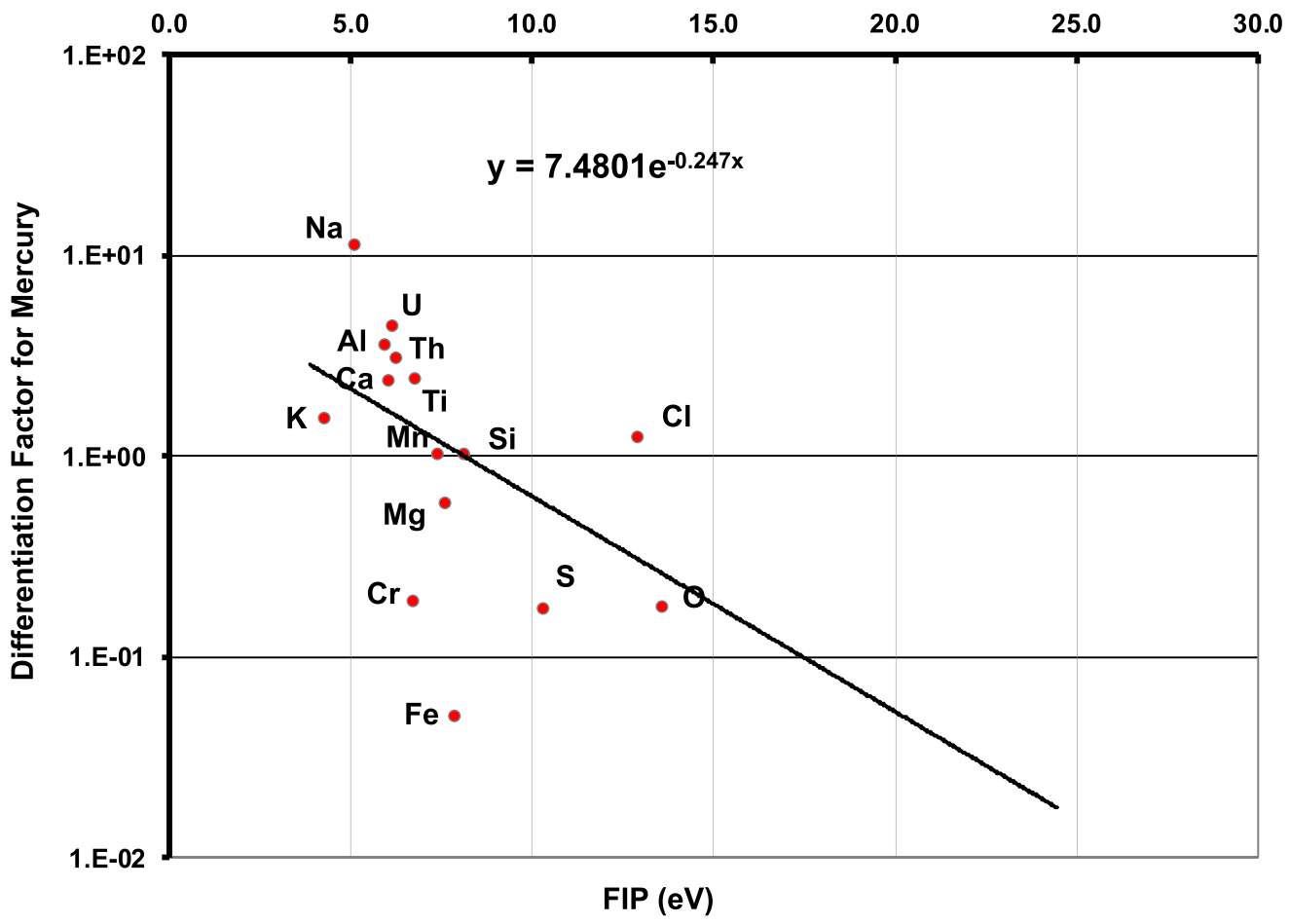

**Figure A7.** Correlation between the observed differentiation factors for Mercury surface and FIP in a semilogarithmic plot. The black line is the theoretical differentiation factor for the volume given by Equation (6) with $R_{PS} \approx 1.5\ R_S$, $d = 0.387$ au, and $T_G(d) =\ 47.445$ K (its equation in inset). Above this line lie elements segregated in the surface, and below are elements depleted from the surface.

## *A.4. Correlations of Observed Differentiation Factors with FIP for Chondrites, Terrestrial Planets, and the Moon*

Figures A6–A10 present as semilogarithmic plots the correlations between observed differentiation factors and FIP of elements for chondrites, Mercury, Venus, Mars, and the

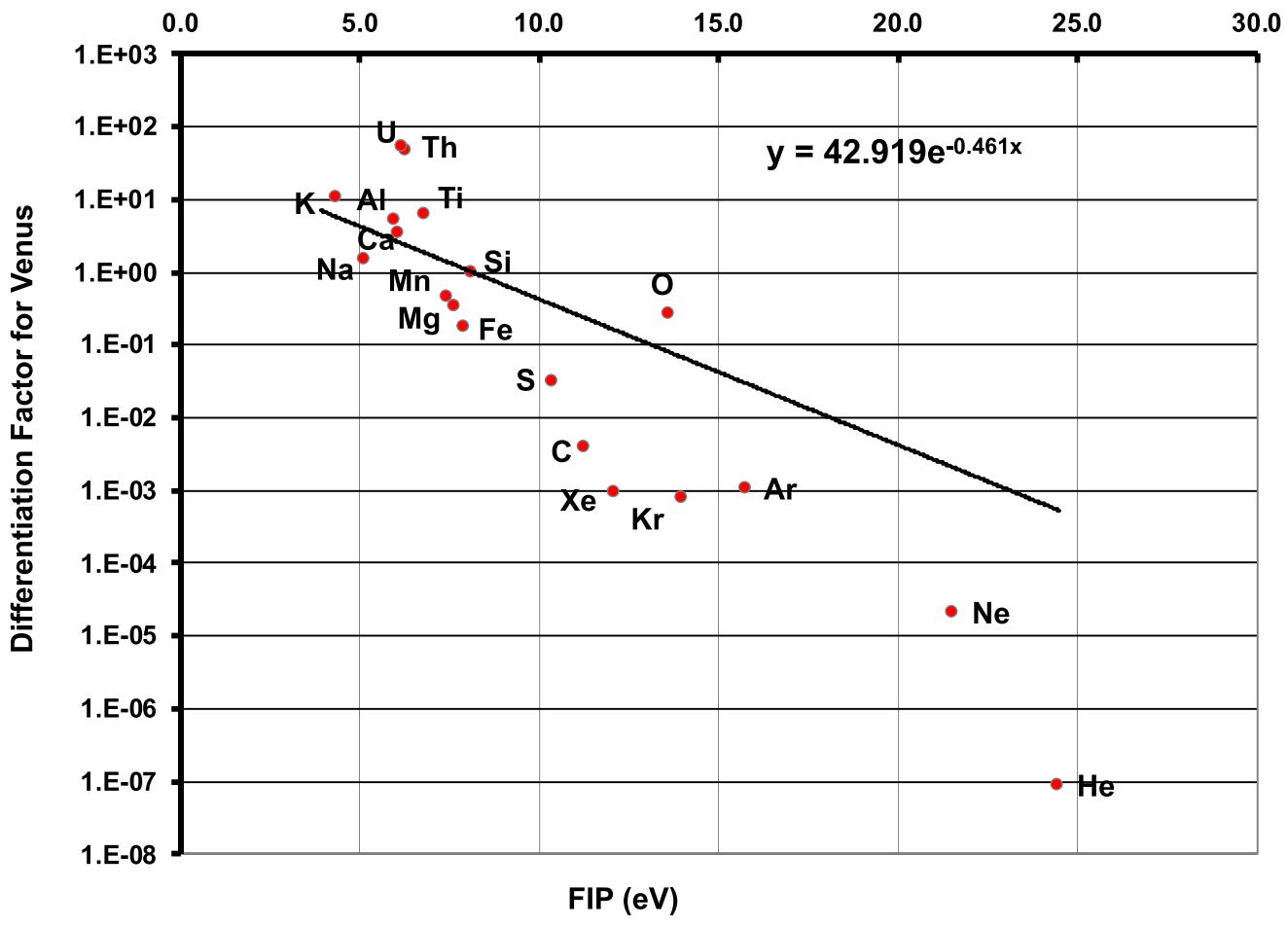

**Figure A8.** Correlation between the observed differentiation factors for Venus and FIP in a semilogarithmic plot. The black line is the theoretical differentiation factor given by Equation (6) with $R_{PS} \approx 1.5 R_S$ $d = 0.723$ au and $T_G(d) =$ 7.276 K (its equation in inset). Above this line lie elements segregated in the surface and below elements depleted from the surface.

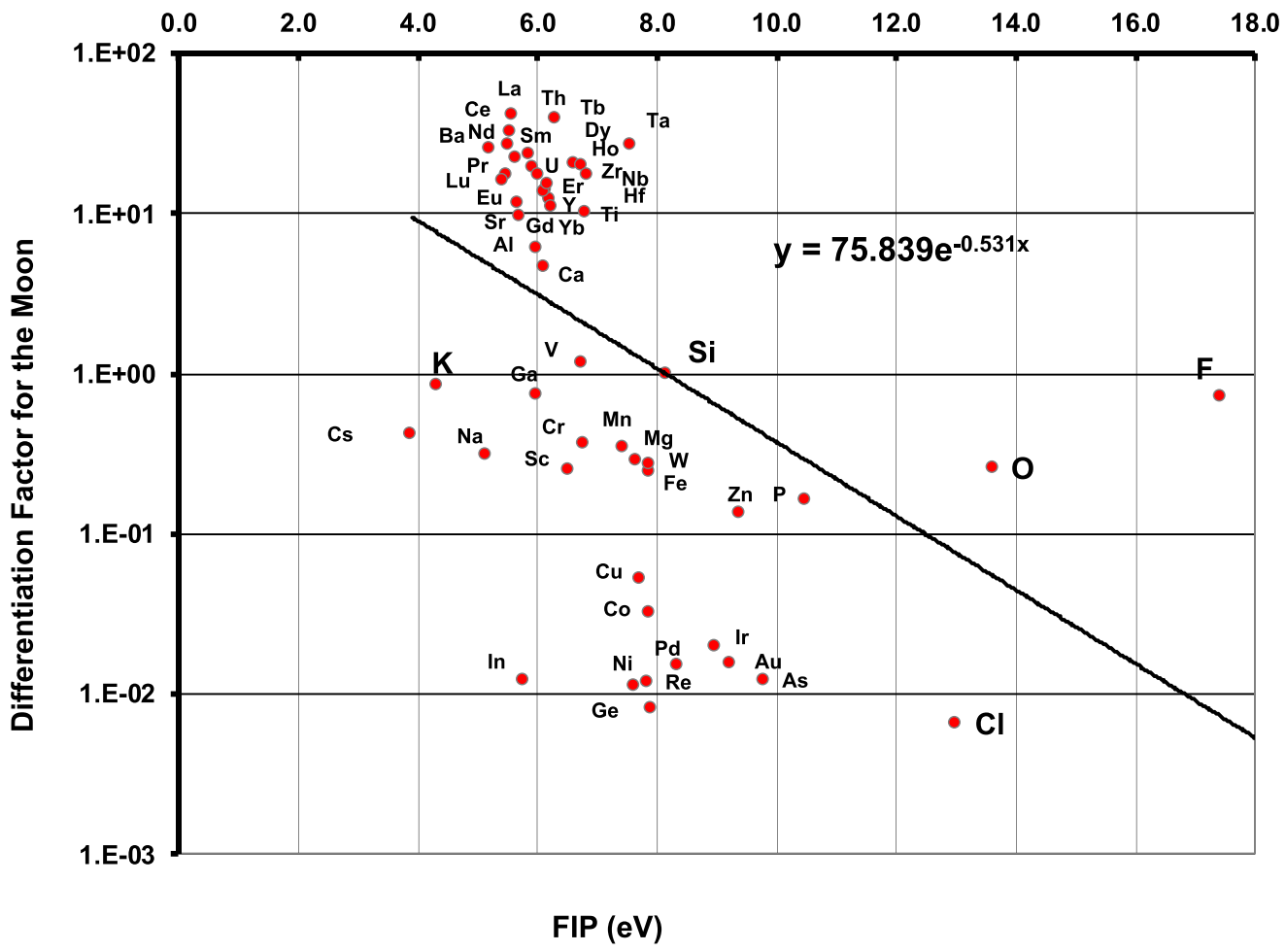

**Figure A10.** Correlation between the observed differentiation factors for the Moon and FIP in a semilogarithmic plot. The black line is the theoretical differentiation factor given by Equation (6) with $R_{PS} \approx 1.5 R_S$ $d = 1.096$ au and $T_G(d) =$ $T_{CB}$ (its equation in inset). Above this line lie elements segregated in the surface, and below are elements depleted from the surface.

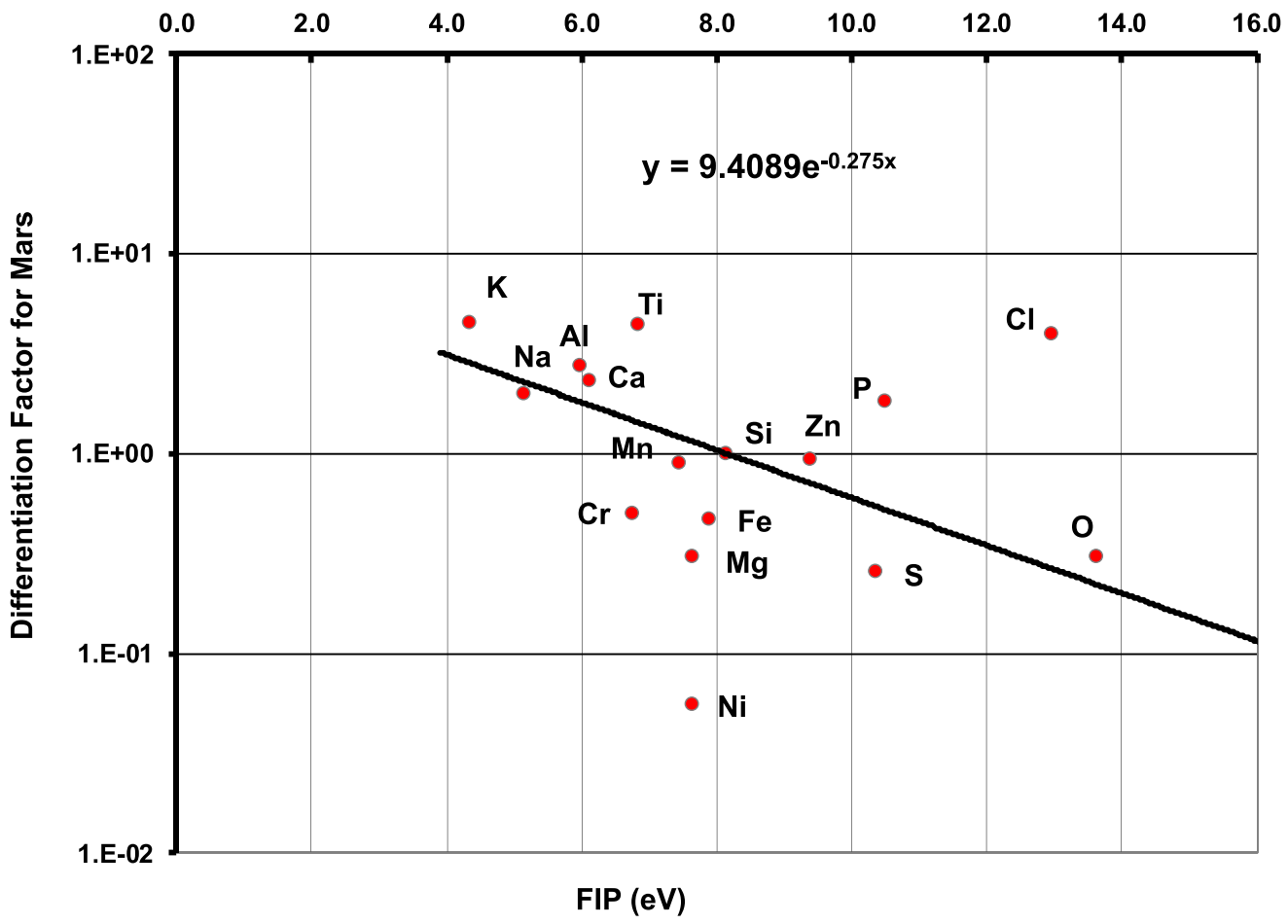

**Figure A9.** Correlation between the observed differentiation factors for Mars and FIP in a semilogarithmic plot. The black line is the theoretical differentiation factor given by Equation (6) with $R_{PS} \approx 1.5 R_S$ $d = 1.523$ au and $T_G(d) =$ $T_{CB}$ (its equation in inset). Above this line lie elements segregated in the surface, and below elements depleted from the surface.

Moon respectively. Chondrites are assumed representative of the asteroid belt.

### A.5. Supplementary Tables

Table A1 provides for major and minor elements the predicted overall composition of the Earth, in weight %, initially (first column), assuming all initial H has escaped (second column), and 4.05 % have been retained internally has hydrides (third column). Table A2 provides the data for Earth and Sun, and Table A3 for Mercury, Venus, the Moon, and Mars, used to create Figures 1 to 10, as well as A6 to A10.

**Table A1**
Predicted Overall Compositions of the Earth in wt% for Major and Minor Elements

| Element | Initially wt% | H Escaped wt% | H Retained wt% |
|---|---|---|---|
| **H** | **83.28** | **0** | **4.050** |
| **C** | **1.098** | **6.579** | **6.323** |
| N | 0.046 | **0.275** | **0.264** |
| **O** | **0.701** | **4.200** | **4.037** |
| **Na** | **0.858** | **5.141** | **4.941** |
| **Mg** | **3.214** | **19.249** | **18.499** |
| **Al** | **0.864** | **5.176** | **4.974** |
| **Si** | **2.622** | **15.702** | **15.091** |
| **S** | **0.308** | **1.845** | **1.773** |
| **K** | **0.161** | **0.966** | **0.929** |
| **Ca** | **0.837** | **5.015** | **4.820** |
| Ti | 0.026 | **0.158** | **0.152** |
| V | 0.0034 | 0.020 | 0.0196 |
| **Cr** | **0.156** | **0.935** | **0.898** |
| Mn | 0.076 | **0.455** | **0.437** |
| **Fe** | **5.310** | **31.804** | **30.566** |
| Co | 0.015 | 0.093 | 0.089 |
| **Ni** | **0.374** | **2.241** | **2.153** |

**Note.** Major elements are given in bold (mass fraction larger than 0.1%: initially; assuming all H has escaped from the Jeans effect; and assuming H is retained as hydrides in lower mantle and core conditions.

**Table A2**
Data (Earth and Sun) Used to Create Figures 1–10

| Element | FIP (eV) | $X$(ES)/$X$(Si) | $X$(SS)/$X$(Si) | $f$(Earth/Sun) Exp. | $f$(Earth/Sun) Theor. | Ln[$X$(ES)/$X$(EV)] | $X$(ES) ppm wt | $X$(EV) ppm wt |
|---|---|---|---|---|---|---|---|---|
| **H** | **13.595** | **2.483E−02** | **1.059E+03** | **2.345E−05** | **2.999E−02** | **−6.122** | **7.00E+03** | **8.33E+05** |
| **He** | **24.481** | **2.837E−08** | **3.539E+02** | **8.016E−11** | **2.709E−05** | **−11.699** | **8.00E−03** | **2.51E+02** |
| **Li** | **5.39** | **7.092E−05** | **1.390E−05** | **5.103E+00** | **5.910E+00** | **0.885** | **2.00E+01** | **2.15E+00** |
| **Be** | **9.32** | **9.929E−06** | **1.979E−07** | **5.018E+01** | **4.705E−01** | **5.701** | **2.80E+00** | **2.44E−03** |
| **B** | **8.296** | **3.546E−05** | **7.335E−06** | **4.835E+00** | **9.097E−01** | **2.702** | **1.00E+01** | **1.75E−01** |
| **C** | **11.256** | **7.092E−04** | **3.098E+00** | **2.289E−04** | **1.352E−01** | **−5.350** | **2.00E+02** | **1.10E+04** |
| **N** | **14.53** | **6.738E−05** | **1.066E+00** | **6.319E−05** | **1.643E−02** | **−4.529** | **1.90E+01** | **4.59E+02** |
| **O** | **13.614** | **1.789E+00** | **9.029E+00** | **1.981E−01** | **2.963E−02** | **2.932** | **5.04E+05** | **7.01E+03** |
| **F** | **17.418** | **2.074E−03** | **5.498E−04** | **3.773E+00** | **2.558E−03** | **8.328** | **5.85E+02** | **3.69E−02** |
| **Ne** | **21.559** | **1.773E−08** | **2.379E+00** | **7.452E−09** | **1.778E−04** | **−9.048** | **5.00E−03** | **1.11E+01** |
| **Na** | **5.138** | **8.567E−02** | **4.710E−02** | **1.819E+00** | **6.951E+00** | **−0.309** | **2.42E+04** | **8.58E+03** |
| **Mg** | **7.644** | **1.063E−01** | **8.856E−01** | **1.201E−01** | **1.384E+00** | **−1.413** | **3.00E+04** | **3.21E+04** |
| **Al** | **5.984** | **2.918E−01** | **8.177E−02** | **3.569E+00** | **4.031E+00** | **0.910** | **8.23E+04** | **8.64E+03** |
| **Si** | **8.149** | **1.000E+00** | **1.000E+00** | **1.000E+00** | **1.000E+00** | **1.032** | **2.82E+05** | **2.62E+04** |
| **P** | **10.484** | **3.723E−03** | **9.173E−03** | **4.059E−01** | **2.223E−01** | **1.634** | **1.05E+03** | **5.35E+01** |
| **S** | **10.357** | **1.241E−03** | **4.870E−01** | **2.548E−03** | **2.413E−01** | **−3.519** | **3.50E+02** | **3.08E+03** |
| **Cl** | **13.01** | **4.076E−03** | **6.625E−03** | **6.153E−01** | **4.371E−02** | **3.676** | **1.15E+03** | **7.59E+00** |
| **Ar** | **15.755** | **4.965E−08** | **1.206E−01** | **4.118E−07** | **7.464E−03** | **−8.773** | **1.40E−02** | **2.36E+01** |
| **K** | **4.339** | **7.411E−02** | **5.293E−03** | **1.400E+01** | **1.163E+01** | **1.218** | **2.09E+04** | **1.61E+03** |
| **Ca** | **6.111** | **1.472E−01** | **8.599E−02** | **1.712E+00** | **3.715E+00** | **0.257** | **4.15E+04** | **8.37E+03** |
| **Sc** | **6.54** | **7.801E−05** | **5.550E−05** | **1.406E+00** | **2.818E+00** | **0.336** | **2.20E+01** | **4.10E+00** |
| **Ti** | **6.82** | **2.004E−02** | **4.281E−03** | **4.680E+00** | **2.353E+00** | **1.719** | **5.65E+03** | **2.64E+02** |
| **V** | **6.74** | **4.255E−04** | **5.231E−04** | **8.135E−01** | **2.478E+00** | **−0.082** | **1.20E+02** | **3.40E+01** |
| **Cr** | **6.764** | **3.617E−04** | **2.441E−02** | **1.482E−02** | **2.440E+00** | **−4.072** | **1.02E+02** | **1.56E+03** |
| **Mn** | **7.432** | **3.369E−03** | **1.826E−02** | **1.845E−01** | **1.587E+00** | **−1.120** | **9.50E+02** | **7.59E+02** |
| **Fe** | **7.87** | **1.996E−01** | **1.692E+00** | **1.180E−01** | **1.197E+00** | **−1.285** | **5.63E+04** | **5.31E+04** |
| **Co** | **7.86** | **8.865E−05** | **4.919E−03** | **1.802E−02** | **1.205E+00** | **−3.171** | **2.50E+01** | **1.55E+02** |
| **Ni** | **7.633** | **2.979E−04** | **1.024E−01** | **2.910E−03** | **1.394E+00** | **−5.140** | **8.40E+01** | **3.74E+03** |
| **Cu** | **7.72638** | **2.128E−04** | **1.243E−03** | **1.711E−01** | **1.313E+00** | **−1.006** | **6.00E+01** | **4.28E+01** |
| **Zn** | **9.3942** | **2.482E−04** | **3.069E−03** | **8.089E−02** | **4.485E−01** | **−0.681** | **7.00E+01** | **3.61E+01** |
| **Ga** | **5.9993** | **6.738E−05** | **9.223E−05** | **7.305E−01** | **3.992E+00** | **−0.667** | **1.90E+01** | **9.65E+00** |
| **Ge** | **7.8994** | **5.319E−06** | **2.970E−04** | **1.791E−02** | **1.174E+00** | **−3.151** | **1.50E+00** | **9.14E+00** |
| **As** | **9.7886** | **6.383E−06** | **1.645E−05** | **3.881E−01** | **3.479E−01** | **1.141** | **1.80E+00** | **1.50E−01** |
| **Se** | **9.75238** | **1.773E−07** | **1.901E−04** | **9.328E−04** | **3.561E−01** | **−4.913** | **5.00E−02** | **1.77E+00** |
| **Br** | **11.81381** | **8.511E−06** | **3.049E−05** | **2.792E−01** | **9.444E−02** | **2.116** | **2.40E+00** | **7.55E−02** |
| **Kr** | **13.99961** | **3.546E−10** | **1.678E−04** | **2.113E−06** | **2.311E−02** | **−8.268** | **1.00E−04** | **1.02E−01** |
| **Rb** | **4.17713** | **3.191E−04** | **2.154E−05** | **1.481E+01** | **1.290E+01** | **1.170** | **9.00E+01** | **7.29E+00** |
| **Sr** | **5.6949** | **1.064E−03** | **7.314E−05** | **1.455E+01** | **4.856E+00** | **2.129** | **3.00E+02** | **9.31E+00** |
| **Y** | **6.2171** | **1.170E−04** | **1.481E−05** | **7.903E+00** | **3.469E+00** | **1.855** | **3.30E+01** | **1.35E+00** |

**Table A2**
(Continued)

| Element | FIP (eV) | $X$(ES)/$X$(Si) | $X$(SS)/$X$(Si) | $f$(Earth/Sun) Exp. | $f$(Earth/Sun) Theor. | Ln[$X$(ES)/$X$(EV)] | $X$(ES) ppm wt | $X$(EV) ppm wt |
|---|---|---|---|---|---|---|---|---|
| **Zr** | **6.6339** | **5.851E−04** | **3.561E−05** | **1.643E+01** | **2.653E+00** | **2.855** | **1.65E+02** | **2.48E+00** |
| **Nb** | **6.75885** | **7.092E−05** | **2.568E−06** | **2.762E+01** | **2.448E+00** | **3.455** | **2.00E+01** | **1.65E−01** |
| **Mo** | **7.09243** | **4.255E−06** | **8.782E−06** | **4.845E−01** | **1.975E+00** | **−0.373** | **1.20E+00** | **4.55E−01** |
| **Tc** | **7.28** | | | | **1.750E+00** | | | |
| **Ru** | **7.3605** | **3.546E−09** | **1.337E−06** | **2.652E−03** | **1.662E+00** | **−5.408** | **1.00E−03** | **5.82E−02** |
| **Rh** | **7.4589** | **3.546E−09** | **5.058E−06** | **7.011E−04** | **1.560E+00** | **−6.676** | **1.00E−03** | **2.07E−01** |
| **Pd** | **8.3369** | **5.319E−08** | **1.856E−06** | **2.866E−02** | **8.860E−01** | **−2.400** | **1.50E−02** | **4.31E−02** |
| **Ag** | **7.5762** | **2.660E−07** | **6.087E−06** | **4.369E−02** | **1.446E+00** | **−2.468** | **7.50E−02** | **2.31E−01** |
| **Cd** | **8.9938** | **5.319E−07** | **7.118E−07** | **7.473E−01** | **5.804E−01** | **1.284** | **1.50E−01** | **1.08E−02** |
| **In** | **5.78636** | **8.865E−07** | **1.484E−05** | **5.973E−02** | **4.578E+00** | **−3.308** | **2.50E−01** | **1.78E+00** |
| **Sn** | **7.3439** | **8.156E−06** | **1.337E−06** | **6.102E+00** | **1.679E+00** | **2.322** | **2.30E+00** | **5.88E−02** |
| **Sb** | **8.6084** | **7.092E−07** | **2.028E−05** | **3.498E−02** | **7.439E−01** | **−2.026** | **2.00E−01** | **3.95E−01** |
| **Te** | **9.0096** | **3.546E−09** | **4.982E−06** | | **5.746E−01** | | **1.00E−03** | |
| **I** | **10.45126** | **1.596E−06** | **2.483E−05** | **6.426E−02** | **2.271E−01** | **−0.231** | **4.50E−01** | **1.48E−01** |
| **Xe** | **12.1298** | **1.064E−10** | **1.737E−06** | **6.125E−05** | **7.705E−02** | **−6.106** | **3.00E−05** | **3.51E−03** |
| **Cs** | **3.8939** | **1.064E−05** | **2.114E−05** | **5.033E−01** | **1.549E+01** | **−2.395** | **3.00E+00** | **8.58E+00** |
| **Ba** | **5.2117** | **1.507E−03** | **2.235E−06** | **6.743E+02** | **6.629E+00** | **5.654** | **4.25E+02** | **3.88E−01** |
| **La** | **5.5769** | **1.383E−04** | **5.811E−06** | **2.380E+01** | **5.239E+00** | **2.545** | **3.90E+01** | **7.98E−01** |
| **Ce** | **5.5387** | **2.358E−04** | **8.670E−07** | **2.720E+02** | **5.370E+00** | **4.957** | **6.65E+01** | **1.22E−01** |
| **Pr** | **5.473** | **4.610E−05** | **4.370E−06** | **1.055E+01** | **5.602E+00** | **1.665** | **1.30E+01** | **6.42E−01** |
| **Nd** | **5.525** | **1.472E−04** | **1.516E−07** | **9.709E+02** | **5.418E+00** | **6.220** | **4.15E+01** | **2.15E−02** |
| **Pm** | **5.582** | | | | **5.222E+00** | | | |
| **Sm** | **5.6436** | **2.500E−05** | **5.354E−07** | **4.670E+01** | **5.019E+00** | **3.262** | **7.05E+00** | **7.04E−02** |
| **Eu** | **5.6704** | **7.092E−06** | **1.965E−06** | **3.610E+00** | **4.933E+00** | **0.719** | **2.00E+00** | **2.54E−01** |
| **Gd** | **6.1501** | **2.199E−05** | **3.615E−07** | **6.082E+01** | **3.622E+00** | **3.852** | **6.20E+00** | **3.43E−02** |
| **Tb** | **5.8638** | **4.255E−06** | **2.305E−06** | **1.846E+00** | **4.356E+00** | **0.173** | **1.20E+00** | **2.63E−01** |
| **Dy** | **5.9389** | **1.844E−05** | **5.277E−07** | **3.494E+01** | **4.150E+00** | **3.162** | **5.20E+00** | **5.74E−02** |
| **Ho** | **6.0215** | **4.610E−06** | **1.545E−06** | **2.985E+00** | **3.935E+00** | **0.755** | **1.30E+00** | **1.59E−01** |
| **Er** | **6.1077** | **1.241E−05** | **2.426E−07** | **5.116E+01** | **3.723E+00** | **3.652** | **3.50E+00** | **2.37E−02** |
| **Tm** | **6.18431** | **1.844E−06** | **1.546E−06** | **1.193E+00** | **3.543E+00** | **−0.057** | **5.20E−01** | **1.44E−01** |
| **Yb** | **6.25416** | **1.135E−05** | **2.343E−07** | **4.844E+01** | **3.388E+00** | **3.692** | **3.20E+00** | **2.08E−02** |
| **Lu** | **5.4259** | **2.837E−06** | **9.874E−07** | **2.873E+00** | **5.775E+00** | **0.334** | **8.00E−01** | **1.49E−01** |
| **Hf** | **6.82507** | **1.064E−05** | **1.359E−07** | **7.830E+01** | **2.346E+00** | **4.540** | **3.00E+00** | **8.35E−03** |
| **Ta** | **7.5496** | **7.092E−06** | **8.893E−07** | **7.975E+00** | **1.471E+00** | **2.722** | **2.00E+00** | **3.43E−02** |
| **W** | **7.864** | **4.433E−06** | **3.681E−07** | **1.204E+01** | **1.201E+00** | **3.337** | **1.25E+00** | **1.16E−02** |
| **Re** | **7.8335** | **2.482E−09** | **4.587E−06** | **5.412E−04** | **1.225E+00** | **−6.693** | **7.00E−04** | **1.47E−01** |
| **Os** | **8.4382** | **5.319E−09** | **4.579E−06** | **1.162E−03** | **8.301E−01** | **−5.540** | **1.50E−03** | **9.97E−02** |
| **Ir** | **8.967** | **3.546E−09** | **8.817E−06** | **4.022E−04** | **5.905E−01** | **−6.260** | **1.00E−03** | **1.36E−01** |
| **Pt** | **8.9587** | **1.773E−08** | **1.354E−06** | **1.309E−02** | **5.937E−01** | **−2.783** | **5.00E−03** | **2.11E−02** |

**Table A2**
(Continued)

| Element | FIP (eV) | $X$(ES)/$X$(Si) | $X$(SS)/$X$(Si) | $f$(Earth/Sun) Exp. | $f$(Earth/Sun) Theor. | Ln[$X$(ES)/$X$(EV)] | $X$(ES) ppm wt | $X$(EV) ppm wt |
|---|---|---|---|---|---|---|---|---|
| **Au** | **9.2255** | **1.418E−08** | **3.206E−06** | **4.425E−03** | **5.000E−01** | **−3.696** | **4.00E−03** | **4.20E−02** |
| **Hg** | **10.4375** | **3.014E−07** | **1.300E−06** | | **2.291E−01** | | **8.50E−02** | |
| **Tl** | **6.1082** | **3.014E−06** | **2.466E−05** | | **3.721E+00** | | **8.50E−01** | |
| **Pb** | **7.41666** | **4.965E−05** | **1.018E−06** | **4.875E+01** | **1.603E+00** | **4.447** | **1.40E+01** | **4.28E−02** |
| **Bi** | **7.2856** | **3.014E−08** | **2.196E−07** | | **1.744E+00** | | **8.50E−03** | |
| **Po** | **8.417** | **7.092E−16** | | | **8.415E−01** | | **2.00E−10** | |
| **At** | | | | | **1.901E+02** | | | |
| **Rn** | **10.7485** | **1.418E−18** | | | **1.875E−01** | | **4.00E−13** | |
| **Fr** | **4.0727** | | | | **1.380E+01** | | | |
| **Ra** | **5.2784** | **3.191E−12** | | | **6.350E+00** | | **9.00E−07** | |
| **Ac** | **5.17** | **1.950E−15** | | | **6.809E+00** | | **5.50E−10** | |
| **Th** | **6.3067** | **3.404E−05** | **2.438E−07** | **1.396E+02** | **3.275E+00** | **4.784** | **9.60E+00** | **2.09E−02** |
| **Pa** | **5.89** | **4.965E−12** | | | **4.283E+00** | | **1.40E−06** | |
| **U** | **6.19405** | **7.092E−06** | **2.501E−07** | **2.836E+01** | **3.521E+00** | **3.118** | **2.00E+00** | **2.31E−02** |

**Note.** Sources are referenced in the text. Blank spaces are for unknown data. FIP: first ionization potential; $X$(ES) mass fraction for Earth's surface; $X$(SS): mass fraction for the Sun's surface (photosphere); $X$(EV): mass fraction for Earth volume (same as in Table A1); $f$(Earth/Sun): differentiation factor as defined in main text; Ln[$X$(ES)/$X$(EV)]: Neperian logarithm of the partition coefficient between Earth's surface and volume.

**Table A3**
Data for Mercury, Venus, the Moon, Mars, and Chondrites Used to Create Figures 1–10

| Element | Mercury ppm wt | $f$(Merc./Sun) Exp. | Venus ppm wt | $f$(Venus/Sun) Exp. | Moon ppm wt | $f$(Moon/Sun) Exp. | Mars ppm wt | $f$(Mars/Sun) Exp. | $f$(Chond./Sun) Exp. |
|---|---|---|---|---|---|---|---|---|---|
| H | | | | | | | | | 1.91E−04 |
| He | | | 6.00E+00 | 8.72E−08 | | | | | 2.71E−10 |
| Li | | | | | | | | | 1.53E+02 |
| Be | | | | | | | | | 1.81E+00 |
| B | | | | | | | | | 1.20E+00 |
| C | | | 4.00E+02 | 3.82E−03 | | | | | 1.09E−01 |
| N | | | | | | | | | 2.84E−02 |
| O | 3.18E+04 | 1.74E−01 | 4.62E+05 | 2.64E−01 | 4.36E+05 | 2.56E−01 | 4.90E+05 | 3.03E−01 | 5.34E−01 |
| F | | | | | 1.07E+02 | 7.11E−01 | | | 8.03E−01 |
| Ne | | | 7.00E+00 | 2.09E−05 | | | | | 1.91E−09 |
| Na | 2.90E+04 | 1.12E+01 | 1.58E+04 | 1.48E+00 | 3.09E+03 | 3.10E−01 | 1.89E+04 | 2.00E+00 | 9.99E−01 |
| Mg | 1.24E+05 | 5.74E−01 | 6.23E+04 | 3.34E−01 | 5.32E+04 | 2.90E−01 | 5.31E+04 | 3.05E−01 | 1.04E+00 |
| Al | 5.91E+04 | 3.55E+00 | 8.77E+04 | 5.06E+00 | 1.06E+05 | 6.11E+00 | 4.53E+04 | 2.75E+00 | 9.76E−01 |
| Si | 2.50E+05 | 1.00E+00 | 2.17E+05 | 1.00E+00 | 2.12E+05 | 1.00E+00 | 2.01E+05 | 1.00E+00 | 1.00E+00 |
| P | | | | | 3.39E+02 | 1.63E−01 | 3.59E+03 | 1.81E+00 | 8.79E−01 |
| S | 2.27E+04 | 1.72E−01 | 3.34E+03 | 3.01E−02 | | 0.00E+00 | 2.73E+04 | 2.57E−01 | 9.60E−01 |
| Cl | 3.50E+03 | 1.22E+00 | | | 1.61E+01 | 6.60E−03 | 9.20E+03 | 3.97E+00 | 5.74E−01 |
| Ar | | | 3.00E+01 | 1.04E−03 | | | | | 1.58E−07 |
| K | 2.00E+03 | 1.51E+00 | 1.19E+04 | 1.04E+01 | 9.49E+02 | 8.46E−01 | 4.83E+03 | 4.54E+00 | 9.42E−01 |
| Ca | 5.50E+04 | 2.33E+00 | 6.60E+04 | 3.32E+00 | 9.22E+04 | 4.62E+00 | 4.38E+04 | 2.31E+00 | 9.03E−01 |
| Sc | | | | | 3.60E+00 | 2.49E−01 | | | 8.02E−01 |
| Ti | 3.08E+03 | 2.40E+00 | 6.07E+03 | 6.07E+00 | 1.12E+04 | 1.02E+01 | 4.56E+03 | 4.40E+00 | 8.02E−01 |
| V | | | | | 1.30E+02 | 1.17E+00 | | | 1.00E+00 |
| Cr | 2.00E+03 | 1.85E−01 | | | 1.79E+03 | 3.61E−01 | 2.34E+03 | 4.99E−01 | 1.04E+00 |
| Mn | 3.50E+03 | 1.01E+00 | 1.29E+03 | 4.42E−01 | 1.02E+03 | 3.47E−01 | 2.51E+03 | 9.00E−01 | 1.30E+00 |
| Fe | 2.04E+04 | 5.05E−02 | 6.68E+04 | 1.72E−01 | 8.42E+04 | 2.46E−01 | 1.52E+05 | 4.67E−01 | 1.06E+00 |
| Co | | | | | 3.42E+01 | 3.21E−02 | | | 9.36E−01 |
| Ni | | | | | 2.39E+02 | 1.13E−02 | 1.11E+03 | 5.52E−02 | 1.00E+00 |

**Table A3**
(Continued)

| Element | Mercury ppm wt | $f$(Merc./Sun) Exp. | Venus ppm wt | $f$(Venus/Sun) Exp. | Moon ppm wt | $f$(Moon/Sun) Exp. | Mars ppm wt | $f$(Mars/Sun) Exp. | $f$(Chond./Sun) Exp. |
|---|---|---|---|---|---|---|---|---|---|
| Cu | | | | | 1.19E+01 | 5.28E−02 | | | 1.13E+00 |
| Zn | | | | | 8.03E+01 | 1.35E−01 | 5.30E+02 | 9.42E−01 | 1.04E+00 |
| Ga | | | | | 8.42E+00 | 7.31E−01 | | | 1.64E+00 |
| Ge | | | | | 4.80E−01 | 7.99E−03 | | | 1.10E+00 |
| As | | | | | 4.84E−02 | 1.22E−02 | | | 8.70E−01 |
| Se | | | | | | | | | 9.97E−01 |
| Br | | | | | | | | | 1.03E+00 |
| Kr | | | 2.70E−02 | 7.57E−04 | | | | | 4.79E−06 |
| Rb | | | | | | | | | 5.72E−01 |
| Sr | | | | | 1.52E+02 | 9.60E+00 | | | 9.71E−01 |
| Y | | | | | 3.82E+01 | 1.22E+01 | | | 9.70E−01 |
| Zr | | | | | 1.60E+02 | 2.07E+01 | | | 1.02E+00 |
| Nb | | | | | 1.07E+01 | 2.00E+01 | | | 9.92E−01 |
| Mo | | | | | | | | | 1.17E+00 |
| Tc | | | | | | | | | |
| Ru | | | | | | | | | 9.05E−01 |
| Rh | | | | | | | | | 9.50E−01 |
| Pd | | | | | 1.69E−02 | 1.48E−02 | | | 1.03E+00 |
| Ag | | | | | | | | | 1.96E+00 |
| Cd | | | | | | | | | 9.33E−01 |
| In | | | | | 1.10E−02 | 1.21E−02 | | | 1.73E−01 |
| Sn | | | | | | | | | 1.29E+00 |
| Sb | | | | | | | | | 1.38E+00 |
| Te | | | | | | | | | |
| I | | | | | | | | | |
| Xe | | | 5.00E−03 | 9.15E−04 | | | | | 5.62E−05 |
| Cs | | | | | 1.55E−01 | 4.21E−01 | | | |
| Ba | | | | | 1.12E+02 | 2.53E+01 | | | 1.04E+00 |
| La | | | | | 1.68E+01 | 4.13E+01 | | | 1.13E+00 |
| Ce | | | | | 3.76E+01 | 3.24E+01 | | | 1.07E+00 |
| Pr | | | | | 2.72E+00 | 1.73E+01 | | | 1.17E+00 |
| Nd | | | | | 2.64E+01 | 2.66E+01 | | | 9.16E−01 |
| Pm | | | | | | | | | |
| Sm | | | | | 7.00E+00 | 2.19E+01 | | | 9.02E−01 |
| Eu | | | | | 1.27E+00 | 1.16E+01 | | | 9.92E−01 |
| Gd | | | | | 6.30E+00 | 1.40E+01 | | | 8.73E−01 |
| Tb | | | | | 1.54E+00 | 2.34E+01 | | | 1.07E+00 |
| Dy | | | | | 9.50E+00 | 1.95E+01 | | | 9.70E−01 |
| Ho | | | | | 2.09E+00 | 1.72E+01 | | | 9.20E−01 |
| Er | | | | | 4.25E+00 | 1.37E+01 | | | 1.04E+00 |
| Tm | | | | | | | | | 1.28E+00 |
| Yb | | | | | 5.00E+00 | 1.10E+01 | | | 7.16E−01 |
| Lu | | | | | 7.09E−01 | 1.62E+01 | | | 1.08E+00 |
| Hf | | | | | 5.13E+00 | 1.74E+01 | | | 7.77E−01 |
| Ta | | | | | 7.75E−01 | 2.71E+01 | | | |
| W | | | | | 1.38E−01 | 2.68E−01 | | | 3.43E−01 |
| Re | | | | | 8.75E−04 | 1.19E−02 | | | |
| Os | | | | | | | | | 8.29E−01 |
| Ir | | | | | 1.96E−02 | 1.96E−02 | | | 9.32E−01 |
| Pt | | | | | | | | | 8.56E−01 |
| Au | | | | | 6.81E−03 | 1.55E−02 | | | 6.63E−01 |
| Hg | | | | | | | | | |
| Tl | | | | | | | | | |
| Pb | | | | | | | | | 1.13E+00 |
| Bi | | | | | | | | | |
| Po | | | | | | | | | |
| At | | | | | | | | | |
| Rn | | | | | | | | | |
| Fr | | | | | | | | | |

**Table A3**
(Continued)

| Element | Mercury ppm wt | $f$(Merc./Sun) Exp. | Venus ppm wt | $f$(Venus/Sun) Exp. | Moon ppm wt | $f$(Moon/Sun) Exp. | Mars ppm wt | $f$(Mars/Sun) Exp. | $f$(Chond./Sun) Exp. |
|---|---|---|---|---|---|---|---|---|---|
| Ra | | | | | | | | | |
| Ac | | | | | | | | | |
| Th | 2.20E−01 | 3.03E+00 | 2.87E+00 | 4.55E+01 | 2.40E+00 | 3.90E+01 | | | 1.00E+00 |
| Pa | | | | | | | | | |
| U | 9.00E−02 | 4.35E+00 | 9.16E−01 | 5.09E+01 | 2.65E−01 | 1.51E+01 | | | 9.53E−01 |

**Note.** Blank spaces are for unknown data. Sources for these data are referenced in the text.

## Appendix B
## Supplementary Equations

Demonstration of Equation (16):

The demonstration is straightforward; note that for all elements $M$,

$$\frac{X_{ES}}{X_{EV}}(M) = \frac{\left(\frac{X_{ES}(M)}{X_{ES}(Si)}\right)\left(\frac{X_{SS}(M)}{X_{SS}(Si)}\right)}{\left(\frac{X_{EV}(M)}{X_{EV}(Si)}\right)\left(\frac{X_{SS}(M)}{X_{SS}(Si)}\right)}\frac{X_{ES}(Si)}{X_{EV}(Si)} \quad (B1)$$

with indices ES, EV, SS, for Earth surface, Earth volume, and average solar system, respectively. By definition of the differentiation (or enrichment) factors $f_i(M)$ (see main text), it follows that

$$\frac{X_{ES}}{X_{EV}}(M) = \frac{f_{ES}(M)}{f_{EV}(M)}\frac{X_{ES}}{X_{EV}}(Si). \quad (B2)$$

The overall mass balance for Earth requires:

$$\sum_M X_{EV}(M) = 1 \quad (B3)$$

or, using (B2):

$$\sum_M X_{ES}(M)\frac{f_{EV}(M)}{f_{ES}(M)} = \frac{X_{ES}(Si)}{X_{EV}(Si)} \quad (B4)$$

from which one determines:

$$X_{EV}(Si) = \frac{X_{ES}(Si)}{\sum_M X_{ES}(M)\frac{f_{EV}(M)}{f_{ES}(M)}} \quad (B5)$$

or:

$$X_{EV}(Si) = \frac{1}{1 + \sum_{M\neq Si} f_{EV}(M)\left(\frac{X_{SS}(M)}{X_{SS}(Si)}\right)}. \quad (B6)$$

Since $f_{EV}(M)$ is known for all elements from Equation (6), and all relative abundances in the solar system are known from experiment, Equations (B2) and (B6) allow us to determine all average mass fractions for the whole Earth volume, including that of element $H$.

Combining (B2) and (B5):

$$\frac{X_{ES}}{X_{EV}}(M) = \frac{f_{ES}(M)}{f_{EV}(M)}\sum_M X_{ES}(M)\frac{f_{EV}(M)}{f_{ES}(M)}. \quad (B7)$$

Taking the natural logarithm of both sides of Equation (B7), one gets Equation (16):

$$\ln\left[\frac{X_{ES}}{X_{EV}}(M)\right] = [\ln f_{ES}(M) - \ln f_{EV}(M)] + \ln\left[\sum_M X_{ES}(M)\frac{f_{EV}(M)}{f_{ES}(M)}\right].$$

## ORCID iDs

Hervé Toulhoat https://orcid.org/0000-0001-8359-4454
Viacheslav Zgonnik https://orcid.org/0000-0003-0968-1757